\documentclass[letterpaper,twocolumn,10pt]{article}
\usepackage{usenix-2020-09}

\usepackage[bottom]{footmisc}

\usepackage[utf8]{inputenc} %

\usepackage{listings}
\usepackage{balance}
\usepackage{enumitem}

\usepackage{tikz}
\usepackage{xcolor}
\usepackage{color}
\usepackage{pgfplots}
\usepackage{comment}

\usepackage{sistyle}
\SIthousandsep{,}

\usepackage{multirow}
\usepackage{booktabs}
\usepackage[disable]{todonotes}

\usepackage{caption}
\usepackage[position=b]{subcaption}

\usepackage{dblfloatfix} 
\graphicspath{{figs/}}

\usetikzlibrary{
	arrows.meta,
	calc,
	colorbrewer,
	external,  
	fit,
	matrix,
	positioning,
	shapes,
	shapes.misc,
	backgrounds
}

\usepackage{cite}

\usepackage{url}
\usepackage{pifont}
\newcommand{\cmark}{\ding{51}}%
\newcommand{\xmark}{\ding{56}}%

\usepackage{xspace}
\newcommand{\etal}{\textit{et~al.}~}
\newcommand{\eg}{\textit{e.g.,}~}
\newcommand{\ie}{\textit{i.e.,}~}
\newcommand{\cf}{\textit{cf.,}~}

\newcommand{\one}{({\em i})\xspace}
\newcommand{\two}{({\em ii})\xspace}
\newcommand{\three}{({\em iii})\xspace}
\newcommand{\four}{({\em iv})\xspace}

\makeatletter
\renewcommand{\paragraph}[1]{\vspace*{0.03in}\noindent{\bf #1.}\hspace{0.25ex \@plus1ex \@minus.2ex}}
\makeatother

\usepackage[absolute,showboxes]{textpos}

\begin{document}

\setlength{\TPHorizModule}{\paperwidth}
\setlength{\TPVertModule}{\paperheight}
\TPMargin{5pt}
\begin{textblock}{0.8}(0.1,0.02)
	\noindent
	\footnotesize
	\centering
	If you cite this paper, please use the USENIX Security reference:
  R. Hiesgen, M. Nawrocki, A. King, A. Dainotti, T. C. Schmidt, and M. Wählisch.
	2022. Spoki: Unveiling a New Wave of Scanners through a Reactive Network Telescope.
	\emph{In Proceedings of 31st USENIX Security Symposium.}\\
	USENIX Association, Berkeley, CA, USA, 18 pages. https://doi.org/TBA
\end{textblock}

\title{\Large \bf Spoki: Unveiling a New Wave of Scanners through a Reactive Network Telescope}

\author{
{\rm Raphael Hiesgen}\\
HAW Hamburg
\and
{\rm Marcin Nawrocki}\\
Freie Universit\"at Berlin
\and
 {\rm Alistair King}\\
Kentik
\and
 {\rm Alberto Dainotti}\\
CAIDA, UC San Diego\\
Georgia Institute of Technology
\and
 {\rm Thomas C. Schmidt}\\
HAW Hamburg
\and
 {\rm Matthias W\"ahlisch}\\
Freie Universit\"at Berlin
} %

\maketitle

\vspace*{-1.5cm}
\begin{abstract}
Large-scale Internet scans are a common method to identify victims of a specific attack. Stateless scanning like in ZMap has been established as an efficient approach to probing at Internet scale. Stateless scans, however, need a second phase to perform the attack. This remains invisible to network telescopes, which only capture the first incoming packet, and is not observed as a related event by honeypots, either.
In this work, we examine Internet-wide scan traffic through Spoki, a
reactive network telescope operating in real-time that we design and
implement. Spoki responds to asynchronous TCP SYN packets and engages
in TCP handshakes initiated in the second phase of two-phase scans.
Because it is extremely lightweight it scales to large prefixes where
it has the unique opportunity to record the first data sequence submitted
within the TCP handshake ACK.
We analyze two-phase scanners during a three months period using
globally deployed Spoki reactive telescopes as well as flow data sets
from IXPs and ISPs.
We find that a predominant fraction of TCP SYNs on the Internet has irregular characteristics. Our findings also provide a clear signature of today's scans as: (i) highly targeted, (ii) scanning activities notably vary between regional vantage points, and (iii) a significant share originates from malicious sources.
\end{abstract}

\section{Introduction}
\label{sec:intro}

Today's Internet is under constant attack~\cite{apt-bhs-07,jkkrs-mtuam-17}. Host scanning is instrumental to discover vulnerable services, create botnets, and launch cyberattacks. Recent stateless scanning techniques that send out handcrafted TCP SYN packets  without utilizing the operating system's TCP/IP stack have advanced speed, scalability, and outreach of malicious scanners---a prominent example of which has been Mirai~\cite{aabbb-umb-17}. Stateless TCP scanners that are seeking more information than open ports need to return to their potential victims with a regular TCP handshake; in this paper we denote them as \textit{two-phase scanners}.

Network telescopes are valuable instruments to monitor the state of the Internet. They collect Internet background radiation~(IBR), which grants insight into the current threat level of the Internet~\cite{cgcps-omeis-05}.  One of the major sources of IBR are scanners that systematically send probes to regions of the IP address space to identify vulnerable hosts.  Naturally, network telescopes detect such behavior~\cite{dkcpp-assb-15}. Nevertheless,  telescopes are passive measurement instruments and the second phase of two-phase scanners remains invisible to them.

In this work, we extend the measurement approach of network telescopes by a reactive response to TCP SYN packets. With our \textit{reactive} network telescope we are able to measure and analyze the behavior of two-phase scanners in a level of detail previously impossible. %
\setcounter{figure}{1}
\begin{figure*}[!b]
  \begin{subfigure}{.33\textwidth}
    \centering
    \includegraphics[width=\linewidth]{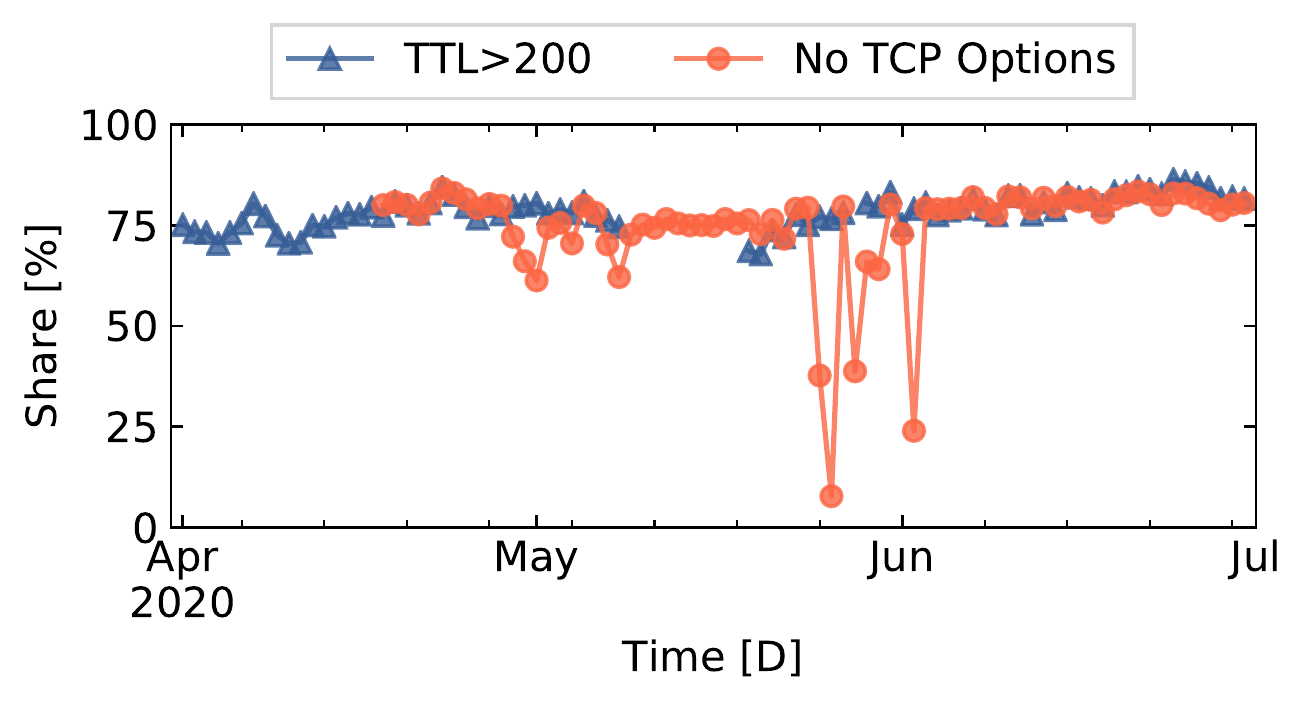}
    \caption{Network telescope data collected in the US.}
    \label{fig:scan:ixp:us}
  \end{subfigure}\hfill
  \begin{subfigure}{.33\textwidth}
    \centering
    \includegraphics[width=\linewidth]{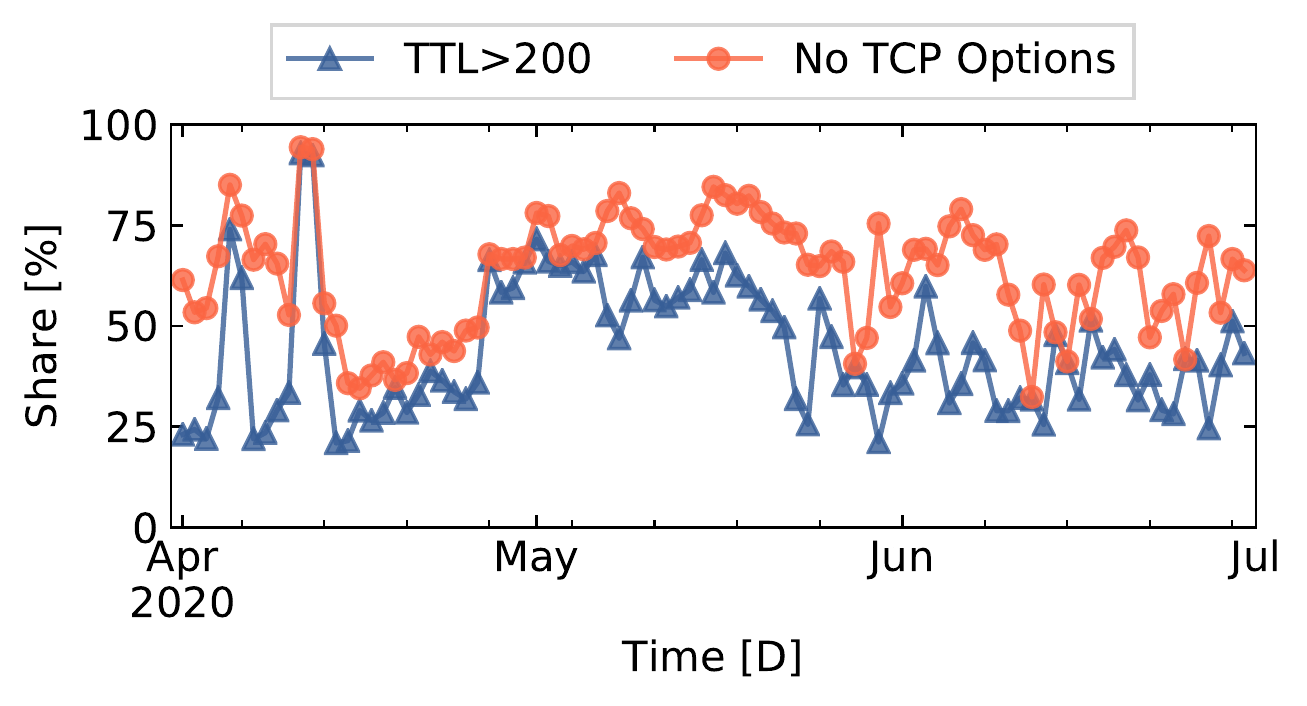}
    \caption{IXP flow data collected in Europe.}
    \label{fig:scan:ixp:eu}
  \end{subfigure}\hfill
  \begin{subfigure}{.33\textwidth}
    \centering
    \includegraphics[width=\linewidth]{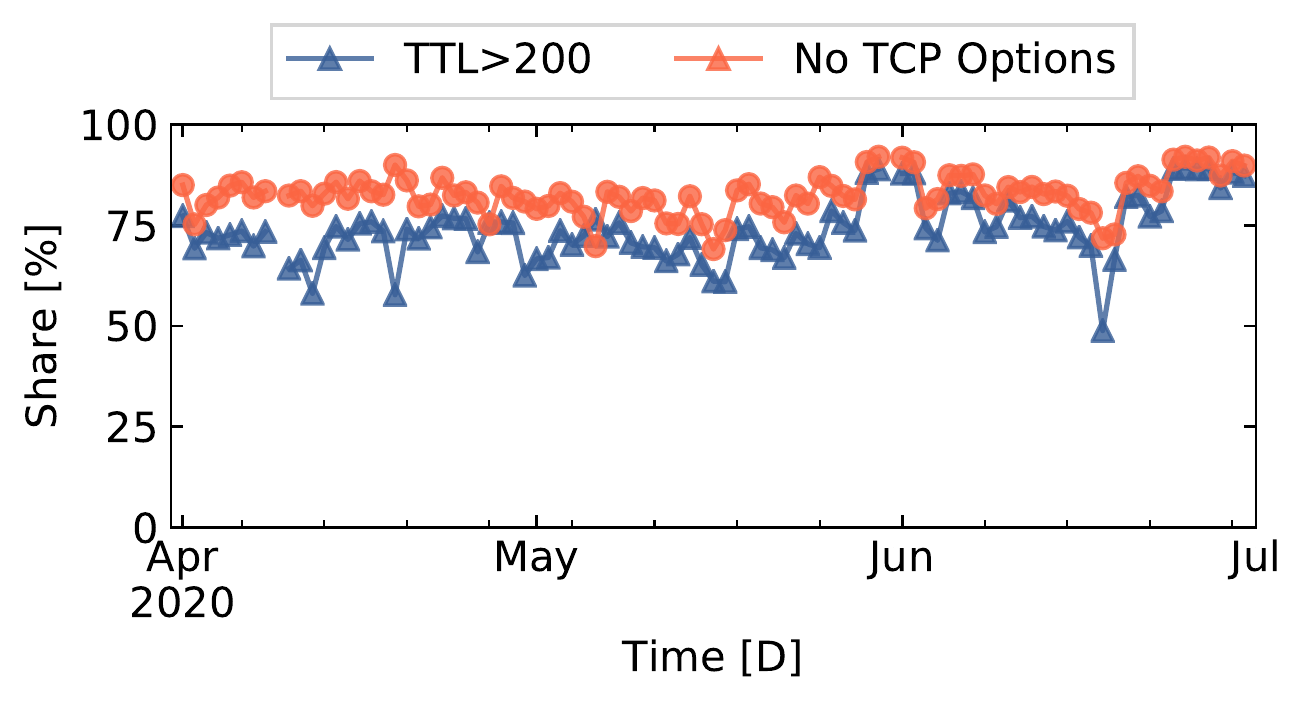}
    \caption{ISP packet captures collected in Asia.}
    \label{fig:scan:ixp:jp}
  \end{subfigure}\hfill
  \caption{Overview of irregular TCP SYN packets at different vantage points, relatively to all TCP SYN packets.}
  \label{fig:scan:ixp}
\end{figure*}
\setcounter{figure}{0}
 Our key contributions and findings are as follows:

\noindent\begin{enumerate}[itemsep=0pt]
\item We design and develop \textbf{Spoki}, a lightweight reactive component that responds in real-time to TCP SYN packets received by the telescope. Spoki 
tries to provoke a second scan phase by interacting with TCP sources and establishing connections when possible.
Since second phases can originate from different ports, we correlate source addresses with timing to match stateless scans to their ``heavyweight'' follow-up probes.
This approach comes with real-time requirements, as 
the client-side state is sensitive to response delays.
Spoki can handle many  connections simultaneously and serve network
telescopes that span /8 IPv4 prefixes.

\item We deploy and leverage a global measurement system that consists of two primary reactive telescopes (in the US and Europe), four control telescopes (US, Europe, Oceania), and two flow collectors from the Internet core (an Asian ISP and a European IXP). We find that \textbf{stateless SYN scanning contributes more than two-thirds of TCP SYN traffic}. %
About a third of these scanning hosts reply to our connection offers and can be attributed to two-phase scanners.

\item We perform an in-depth study of two-phase scanners, including their behavior, service targets, and (unspoofed) origins. We find that the \textbf{activities of two-phase scanners are significantly more targeted than common one-phase scanners}. This includes the contacted services, temporal orchestration, and proximity of sources. When comparing data from our two reactive telescopes, we even see a clear signature of destination-specific service port selection of the scanners.   

\item We leverage Spoki also to capture the first data sequence of the scanner and perform a \textbf{detailed content analysis}. Our findings reveal extensive banner grabbing, but also repeated signatures of well-known exploits in particular of home routers, mobiles, and IoT devices. We are able to monitor shell code activities, as well as injections of code, some of which are unseen in public sources.

\item We measure and analyze localized two-phase scanning. To our surprise, \textbf{scanning activities in Europe largely peak at sources from the same /16 covering IP prefix} as the victims, whereas similar localization patterns are not present in the US or Asia. We verify these results through an extensive flow-data analysis.    
\end{enumerate}

To the best of our knowledge, this work is the first to systematically study two-phase scanners using a reactive network telescope. Our findings shed %
light on a rapidly growing ecosystem of highly performant malicious actors.

\begin{figure}
  \centering
  \includegraphics[width=\linewidth]{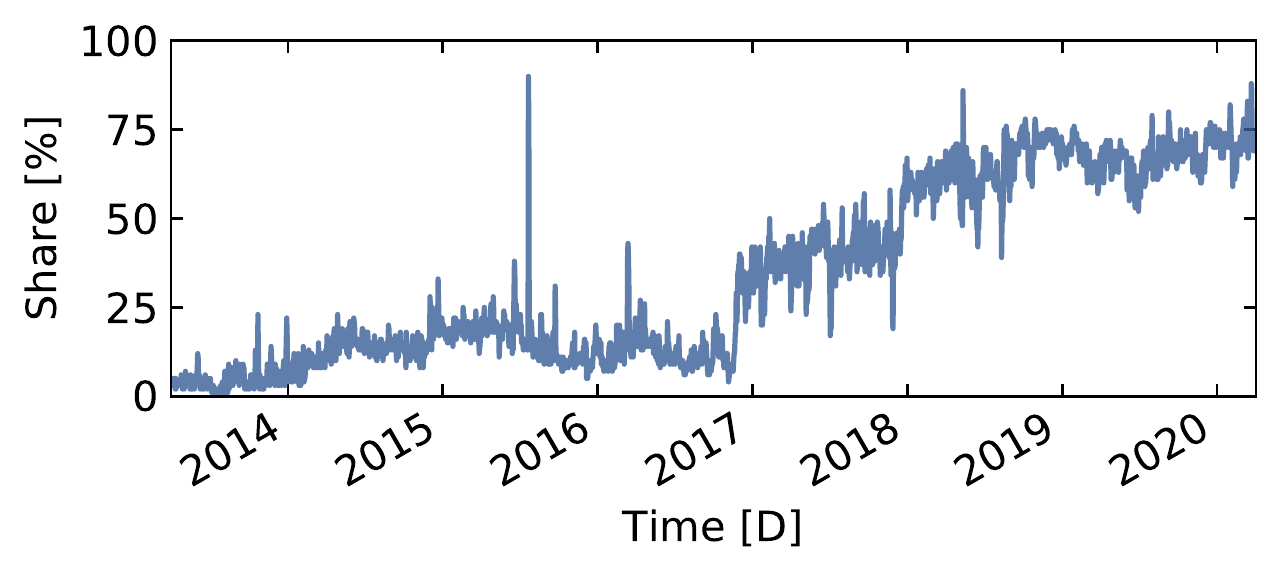}
  \caption{Share of IPv4 packets with a $TTL > 200$ since 2013.}
  \label{fig:scan:ttl200}
\end{figure}
\setcounter{figure}{2}

\section{The Rise of Irregular TCP SYNs}
\label{sec:scanners}

In this section, we present our initial observations that motivate this study.
Specifically, we discuss reasons behind---and quantify the amount of---irregular
TCP/IP packets in traffic measured at different vantage points in the Internet.

\subsection{Regular versus Irregular TCP/IP Packets}

Regular applications use the  TCP/IP stack of an operating system to build and transmit packets.
Few exceptions handcraft packets using raw sockets, mainly to
accelerate scanning~\cite{k-paketto-02,dwh-zfiss-13,lll-iuru-05,g-meim-13} or to explicitly set header field values~\cite{gd-dcusd-20,gbd-ripst-16}.
We consider packets to be \emph{irregular} if their header values
differ from common system stack behavior, \eg by their TTL value.

\autoref{fig:scan:ttl200} plots the growth of IPv4 packets with a TTL larger
than 200 as the share of all packets since 2013, monitored in a large network
telescope which spans the majority of a /8 IPv4 prefix. 
This telescope is used frequently to understand the state of the Internet, \eg \cite{bdcsk-libro-15,dkcpp-assb-15,dbkck-eiasu-14}.
To our surprise, we observe a steep rise over seven years up to the current share of around 80\%.
Such a high share is uncommon in regular traffic, as the prevalent operating systems use lower default values for TCP packets~\cite{s-dtvdo-20}.
Monitoring TTL values as an indicator for irregular traffic is a well-known technique that has been explored in various contexts~\cite{bdcsk-libro-15, jws-hfeda-03}. 

To confirm that irregular packets are not a local phenomenon or specific to this
vantage point, we compare the telescope data with two additional vantage points,
a large, regional European Internet Exchange Point (IXP) and an intercontinental
link of an Asian Internet Service Provider~(ISP). We do not disclose
the specific locations of these vantage points since they belong to commercial
entities.

Among the packets that have high TTL values, we notice a significant portion of TCP packets without TCP options, another irregularity.
The OSes dominant on the Internet offer several options during TCP handshake including ``maximum segment size'', ``SACK permitted'', timestamps, and a window scale.
\autoref{fig:scan:ixp} compares the shares of TCP SYN packets with a TTL value above 200 or without TCP options in relation to all TCP SYN packets.
The shares largely overlap since many packets exhibit both characteristics.
While irregular traffic is not surprising in the telescope data (\autoref{fig:scan:ixp:us}), the share of irregular SYNs in the regular IXP and ISP traffic still rises well above 50\%  and spikes up to 90\% (\autoref{fig:scan:ixp:eu}, \autoref{fig:scan:ixp:jp}).
This phenomenon appears prevalent at all three vantage points.

Irregular SYN packets not only dominate in quantity, they also cover a
much wider port range. While more than 40\% of the \textit{regular}
SYN packets in each telescope focus on Windows vulnerabilities (445,
3389, 1433), \textit{irregular} SYNs try to contact a large variety of
services including SSL, SMTPS, 3074 (Xbox), and gaming ports. 
Further activities narrow down to responding ports in two-phase scans as discussed below. 

Examining the packets in more detail, we notice an IP~ID of 54321 in 5\% of all
irregular packets. This value is well-known to be used by ZMap~\cite{dwh-zfiss-13}, the
first widely deployed \textit{stateless} scanner.
We will now argue that the different traffic features of irregular packets relate to two-phase scans and that this type of scans may have malicious intent.

\subsection{Two-phase Scanning}

By manually analyzing network traces and the open source code of
popular scanning tools, we find that stateless scans use irregular TCP
SYNs as probe packets.
Instead of using the TCP stack of the operating system to issue probes, \textit{stateless scanners}
handcraft and send SYN packets from user space.
Encoding schemes such as SYN cookies~\cite{gbd-ripst-16} are further
used to match response packets to the initial probes.
This lightweight approach eliminates the need for local state, thus abandons memory constraints and speeds up Internet-wide scans.

\begin{figure}%
  \centering
  \includegraphics[width=\linewidth]{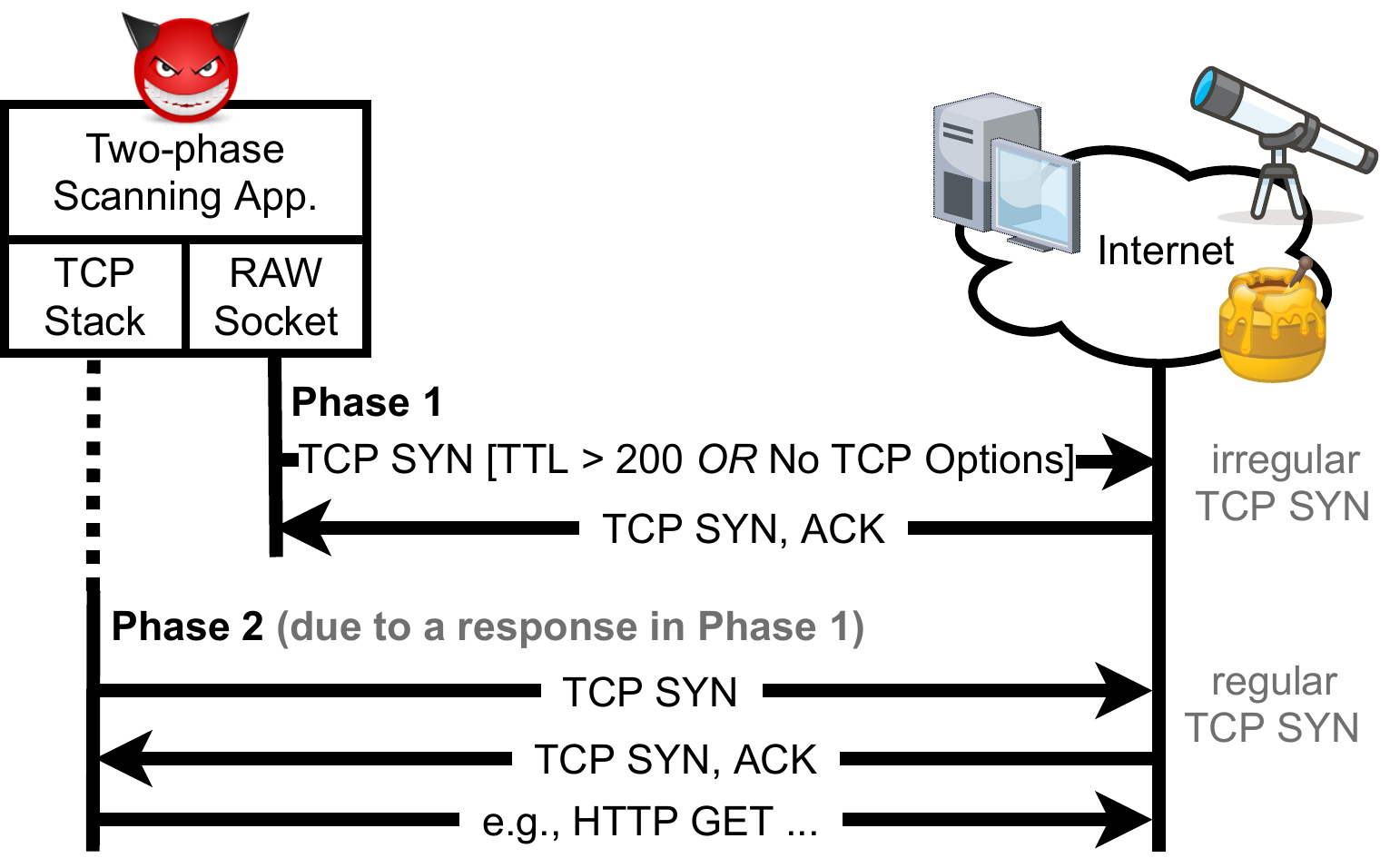}
  \caption{Two-phase scanning.}
  \label{fig:overview-two-phase-scan}\vspace{-7pt}
\end{figure}

Traditional scans  establish a TCP
connection and can then gather information from running
services, \eg through banner grabbing or port-specific payloads. Stateless
scanners often implement this step only in a second (stateful) phase, such that they only
 connect to targets identified as responsive through the (stateless)
first phase (see \autoref{fig:overview-two-phase-scan}). This
\textit{two-phase}
approach drastically
reduces the complexity of scanning and gives both researchers and attackers the
ability to efficiently probe large parts of the IPv4 address space.

Another way of thinking about this scan methodology is by protocol layers.
The first phase tests connectivity on the transport layer before the second phase checks the application layer.
This separation into phases only pays off due to the extra round trip required for TCP to send the application layer payload.
 UDP can check both layers with the first packet and does not need a second phase. Hence, we focus on TCP.
 
The Mirai botnet is known to have used such a multi-phase scan method to find potential
victim hosts using lightweight stateless scans before brute-forcing the login
credentials in a second phase~\cite{aabbb-umb-17}. The infection itself was performed asynchronously once valid login credentials were discovered.

For the purpose of this paper, we denote a scanner as an entity or
program that sends probes to one or more remote host with the goal to
find hosts that react in a certain way to the packets or their payload. Such a behavior could be a specific response message or a triggered action in a local process.

\subsection{Malicious Activities}
\label{sub:mirai}

We now show that irregular SYNs and two-phase scanning are largely used for malicious activities.
An activity is `malicious' if it has the intent to impede a host functionality, accesses it or use its resources without consent of the owner, or perform other harmful activities that involve the system.  We use `attack' for a targeted action with a  malicious purpose.

To this end, we analyze the first significant increase of irregular scans at the end of 2016 (see \autoref{fig:scan:ttl200}) in more detail and hypothesize that these relate to the Mirai botnet~\cite{aabbb-umb-17, wktpl-hs-18}.
We use data from our IXP vantage point because its traffic provides an inter-domain view, covering multiple networks and giving additional insights compared to honeypots~\cite{njsw-fsdat-21}.

Starting on Sep.~2016, Mirai performed several distributed
denial-of-service (DDoS) attacks~\cite{aabbb-umb-17} that led to the
largest attacks recorded to date.
The Mirai method was adopted quickly due to its efficient implementation of stateless scanning and the misuse of insecure default passwords in IoT~products.
At its prime, Mirai controlled 400k~bots.
The public release of the Mirai source code enabled a quick spread of new variants.
These botnets slightly diverge from the original implementation but follow the same two-phase scanning~approach.

\begin{figure}[b]
  \centering
  \includegraphics[width=\linewidth]{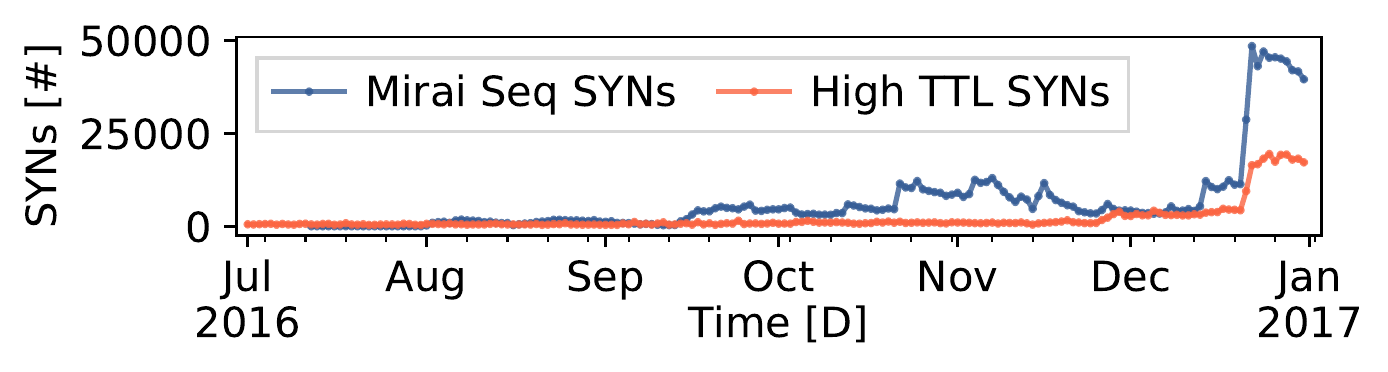}
  \caption{Number of TCP SYNs with Mirai sequence number fingerprint and TCP SYNs with a TTL $> 200$ at the EU~IXP.}
  \label{fig:mirai_num_packets}
\end{figure}

\begin{figure}%
  \centering
  \includegraphics[width=\linewidth]{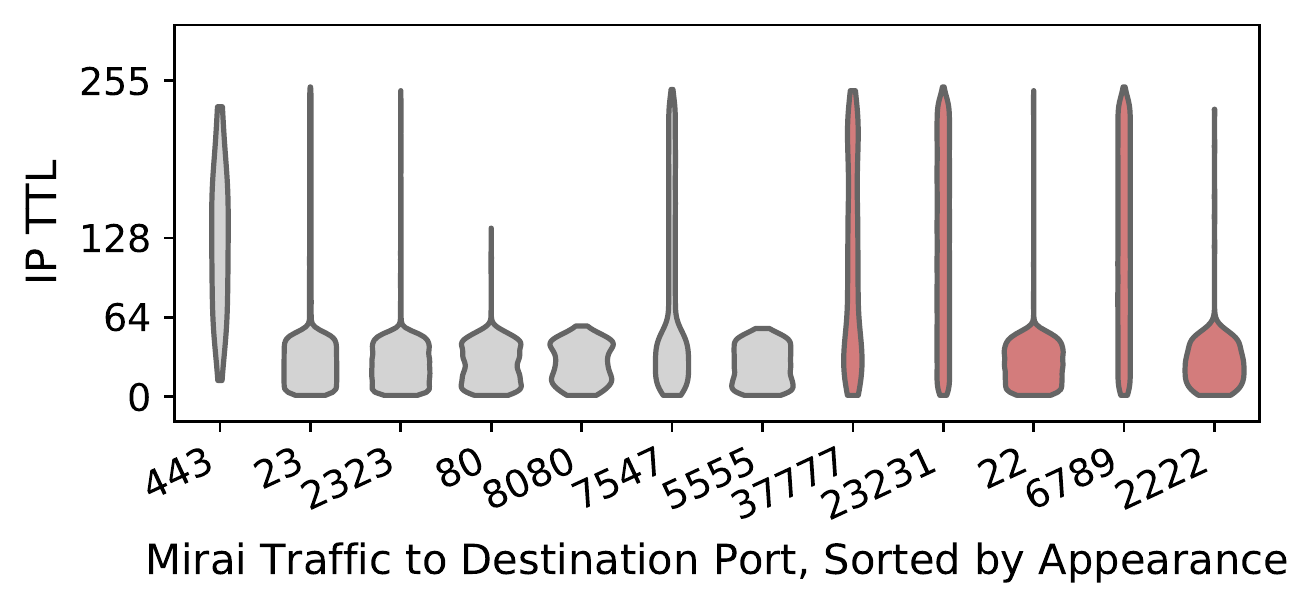}
  \caption{Violin plot of the IP TTL distributions of various Mirai variants. Variants that emerged in Dec.~2016 (marked red) also utilized high TTLs.}
  \label{fig:mirai_violin_ttl}\vspace{-6pt}
\end{figure}

The original Mirai bot sent TCP packets with a sequence number set to the destination IP address instead of a random 32-bit integer.
We observe a clear increase of such packets in mid Sep.~2016 (see \autoref{fig:mirai_num_packets}).
Even though the original Mirai implementation used a fixed TTL
value of 64, later variants utilized higher TTLs as well.
\autoref{fig:mirai_violin_ttl} shows the distribution of TTLs in traffic with the Mirai fingerprint per common target port from Oct.~2016 to Jan.~2017.
This evolution led to the increase of high TTL SYNs at the IXP in
Dec.~2016, consistently with our telescope observations in \autoref{fig:scan:ttl200}.
We also observe an increased number of Mirai-like bots
targeting ports commonly attacked by Mirai, which is well-known to have inspired a variety of successors~\cite{avast-mirai-ongoing} changing fingerprints~\cite{aabbb-umb-17}.
Nearly all scans were initiated by different source IP addresses, and most of the IP addresses belong to different
/24 IP prefixes. This highlights a distributed set of hosts, which is common for distributed botnets but not for research scanners.

To summarize, we conjecture that the Mirai stateless scan engine and its
public source code paved the way for new malware, which incorporated its functionality but avoided the exact same fingerprint.
A steady growth of irregular TCP SYNs on the Internet lets us observe the consequences of Mirai, though. 
Reviewing the rise of packets with high TTLs in \autoref{fig:scan:ttl200} two time points stand out.
The first rise in high TTL packets happened around 2014, which correlates with the release of  ZMap  in 2013~\cite{dwh-zfiss-13}.
A second, higher increase in high TTL packets at the end of 2016 matches the observations in \autoref{fig:mirai_num_packets}.
From the respective shares of high TTL packets among all packets the impact of the Mirai release was bigger than the impact of the ZMap release, which could indicate a strong use of irregular SYNs among malicious actors.
In \autoref{sec:eval:behavior}
we will present more evidence that two-phase scanning is used for malicious activities to date.

\subsection{Revealing Two-phase Scanners}

To further analyze irregular SYN scanning, and in particular learn about two-phase scanner activities,
we built a system that interacts with SYN sources
and actively engages in the second phase (see \autoref{sec:design} for details).
In the remainder of this paper, we mark any TCP SYN packet as \emph{irregular
SYN} if it exhibits one or more of the following three characteristics: \one TTL
$>$ 200, \two no TCP options, \three an IP ID of 54321. 
While this feature set represents our current fingerprint of two-phase scanners, tools can evolve to use different values and evade the present detection scheme. Stateless scanning, however, which comprises the first phase,  requires pre-cached, hand-crafted headers that will be detectable by some eventually altered feature set. Hence, our methodology of fingerprinting two-phase scanners will remain feasible with appropriately changing parameters. 

Just like a regular
host, our system accepts incoming TCP connections and collects payload on
success. Unlike a regular host, our system listens on all IP addresses and ports
of an otherwise unutilized IP prefix\footnote{Both our telescopes use dark IP space that was unused for a long time.}, \ie it acts as a \textit{reactive} network
telescope.

\begin{figure}%
    \centering
    \includegraphics[width=1.06\columnwidth]{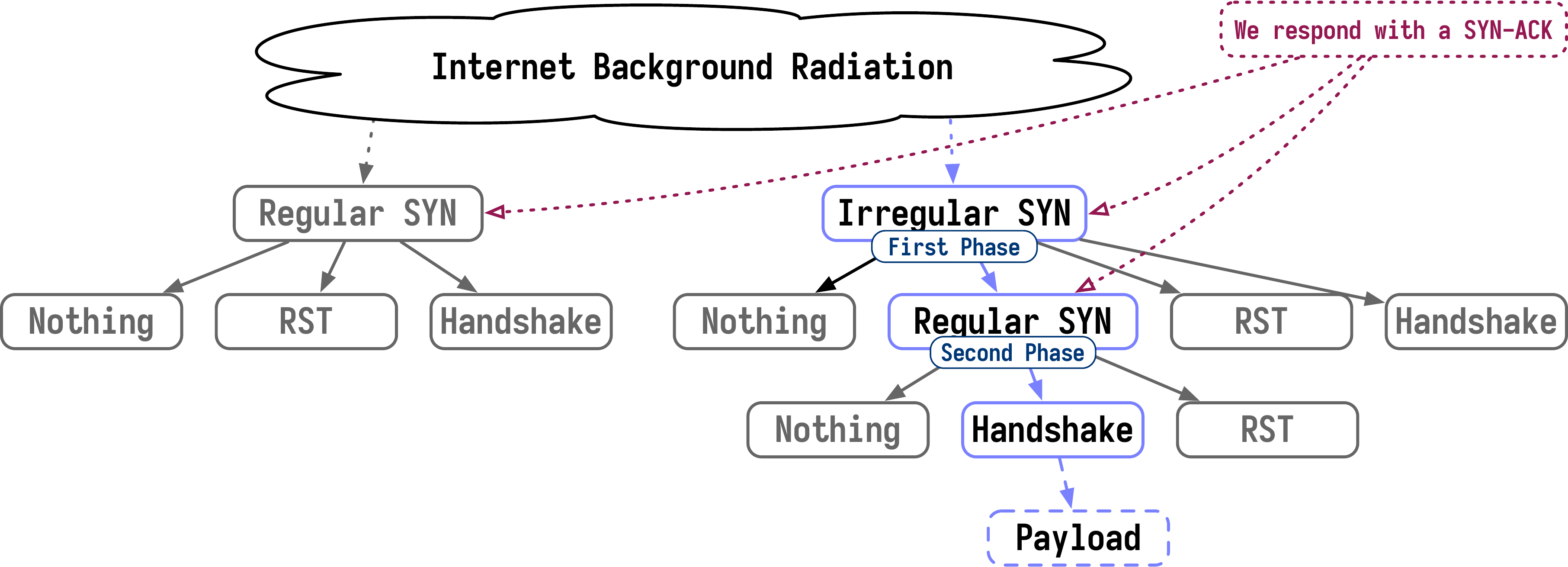}
	\caption{\vspace*{.5cm}Taxonomy of observed scan events.}
    \label{fig:scan:taxonomy}\vspace{-9pt}
\end{figure}

\autoref{fig:scan:taxonomy} depicts a taxonomy of observed scan events. SYN
packets that arrive at our vantage point can be split into regular and irregular
SYNs. We reply with a SYN-ACK in both cases and the remote host either ignores our response, replies with a RST, or completes the TCP~handshake with an ACK. %
In the events we look for, the sender appears to ignore our SYN-ACK response to an irregular SYN, but
shortly after reaches the same IP address and port with a regular SYN in a second
phase (right branch of
our taxonomy in the figure). Once the handshake of the second phase is complete,
we collect the payload and reset the connection after a short delay.

This classification is not possible in traditional telescopes, which do not engage in the second phase. Honeypots can observe both phases, but they normally neither correlate the events nor do they analyze irregularities in received packets. Performing these steps in real-time is a unique feature of our detection system.

A second SYN packet can follow a regular TCP SYN as well. We found one order of magnitude fewer of these cases and they almost never carried content. We were able to attribute 20\% of them to well-known scanning projects such as Shodan, while 75\% are rare events ($\leq$ 4 SYN/month). 
These observations confirm that irregular SYN packets are characteristic for the two-phase events we want to investigate.   

\begin{figure} %
  \centering
  \includegraphics[width=\columnwidth]{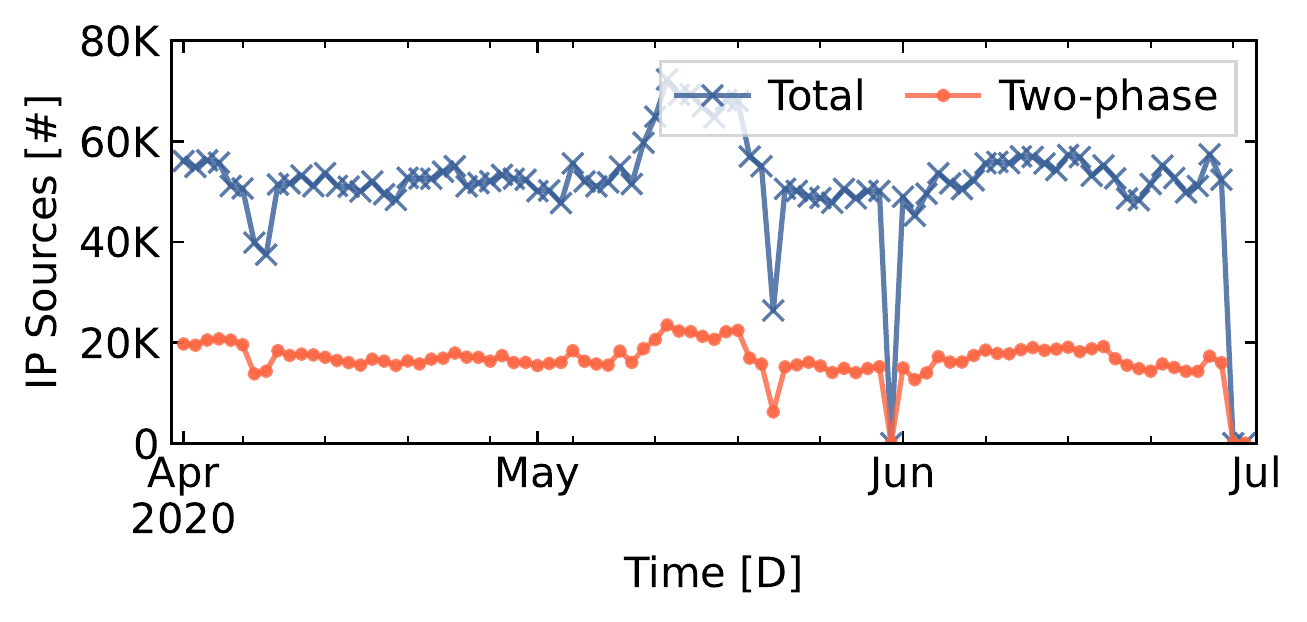}
	\caption{Share of observed two-phase sources among sources that send TCP SYN packets, collected in the US. Very similar results are visible at our EU~telescope (see \autoref{apx:scan:two-phase-sources:eu}).}
  \label{fig:scan:two-phase-sources}
\end{figure}

\autoref{fig:scan:two-phase-sources} compares the total number of source IP
addresses contacts per day with the number of these sources that follow the
two-phase scanner pattern as observed from Apr. through Jun.~2020. 
(We exclude addresses of known scan projects---see \autoref{sec:eval} for details).
Both our vantage points see a consistent share of around 30\% sources that
perform two-phase scans. 
The drops in a few specific days are due to data loss.

Irregular SYNs are significant in telescope (\ie unsolicited)
traffic and hold a noticeable share in flow data of generic traffic (\eg
visible at a large IXP). Our observations show a relatively stable share of about 30\% sources that perform two-phase scans in the telescope. These sources are responsible for a much greater share of packets when also considering regular scanner behavior.
Understanding the behavior of these scanners as well as their intentions is of interest considering their involvement in major security incidents in the past.

\section{Spoki: A Reactive Network Telescope}
\label{sec:design}

We now introduce Spoki, our real-time \emph{reactive} network telescope.
Spoki supports both the analysis of arriving traffic and active measurements to track the behavior of the sources.

Designing Spoki faces the following challenges.
First, Spoki needs to be scalable in terms of traffic volume and resources.
Even if we are only leveraging a couple of unused /24 blocks in this study, our target
for stable deployment is one of the largest IPv4 network telescopes (\ie three quarters of
a /8 prefix) listening to more than 12M~IP~addresses. %
Current traffic peaks include up to 83.4M packets per minute.
Second, Spoki should ensure that scanners receive the SYN ACK in time to engage in the second phase.
This means reacting to any incoming packet in near real-time.
We need to emulate TCP/IP network stack behavior for any of the listening IP
addresses as if each IP~address was hosted on a separate system.

\begin{figure}[b]
  \centering
  \includegraphics[width=\columnwidth]{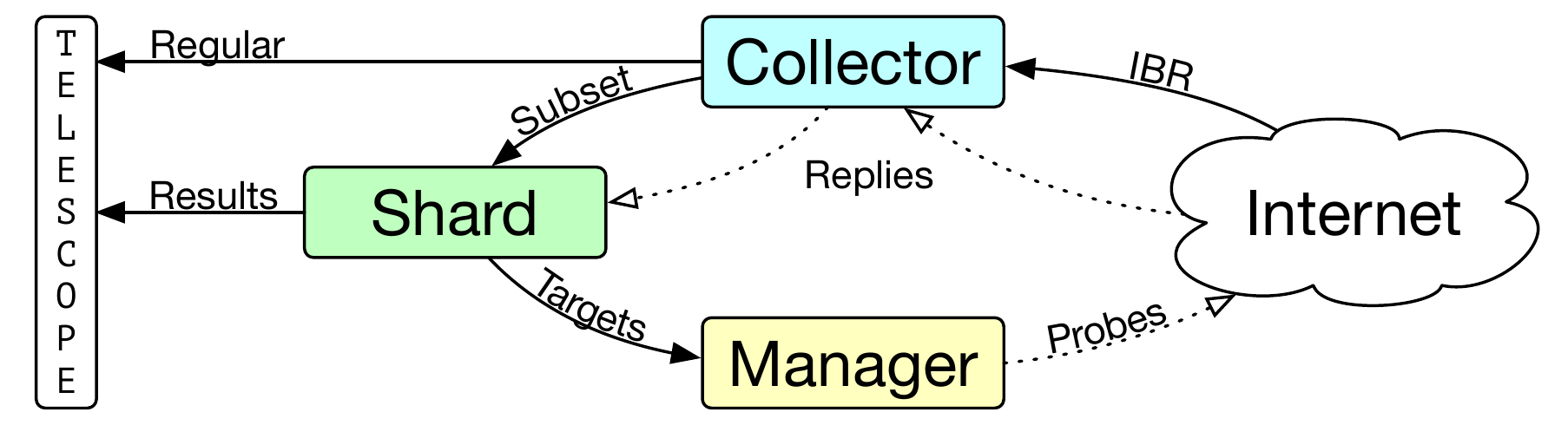}
  \caption{Traffic flow for reactive probing in the telescope.}
  \label{fig:design:traffic_flow}
\end{figure}

The core building blocks of our system are depicted
in \autoref{fig:design:traffic_flow}. IBR
(\ie unsolicited traffic) including irregular SYNs
originating from the whole Internet is captured by a network telescope. Spoki
``subscribes'' to listen to a subset (or all) of the telescope IP~addresses
and extracts the source IP~addresses of TCP SYN packets along with the necessary
information to craft valid replies. It then uses
Scamper~\cite{l-ssepp-10}, a specialized open-source probing application, to
send probes that, to the extracted targets, appear as valid responses (\eg each
 source address of a probe matches the destination address of the original packet
observed and the ACK number is set consistently).
As a result, replies to these probes arrive in the telescope mixed with regular IBR. Spoki identifies replies and stores them for further processing.
We carefully configure Spoki to not operate as an amplifier (see \autoref{apx:ethics-dos}).
For a discussion of related work in the context of Spoki, see \autoref{sec:background}.

\subsection{Implementing Spoki}
\label{sub:design:impl}

Our goal is to design, implement, and deploy a system that can handle the expected traffic in real-time. We leverage the actor programming paradigm.
Actors provide scalable components that pass messages among concurrently running computations and separate our system into smaller, specialized building blocks. 
We implement Spoki in native C++ extended by the C++ Actor Framework (CAF)~\cite{chs-rapc-16}. 

CAF combines the benefits of native program execution with a high level of abstraction.
Transparent messages passing between actors makes the system distributable and allows scaling across multiple cores as well as multiple hosts.
In our case, packets with different source addresses are independent events and can thus be handled concurrently, which makes them a perfect fit for concurrent processing offered by actors.
This foundation allows us to focus on the high-level analysis and data flow
instead of investing effort into fitting Spoki into the specific deployment environment.

\begin{figure}[b]
  \centering
  \includegraphics[width=\columnwidth]{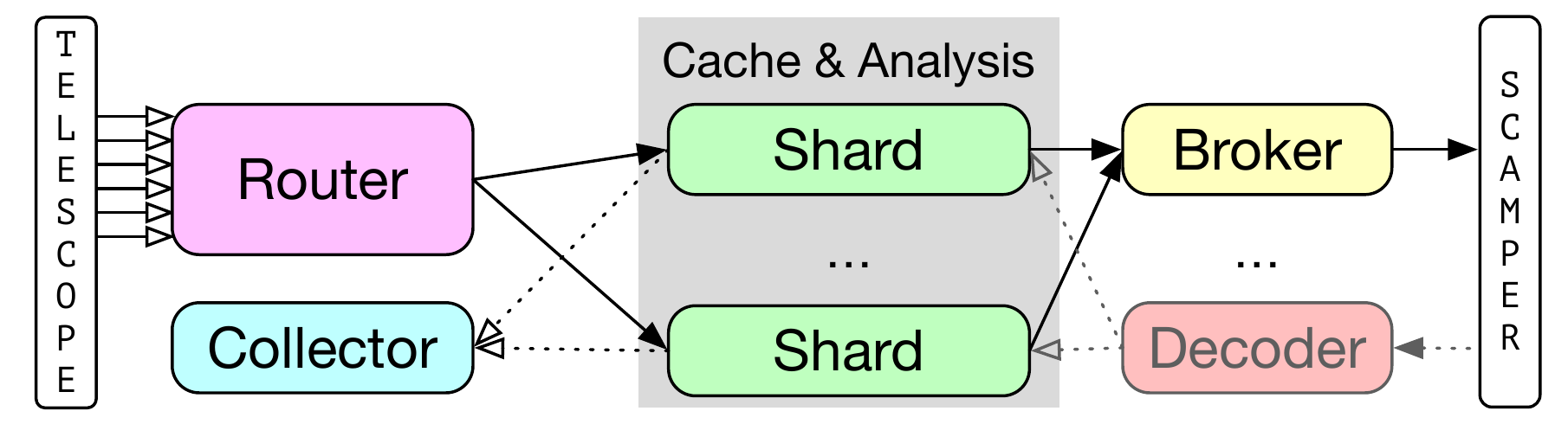}
  \caption{The architecture of Spoki.}
  \label{fig:design:system_arch}
\end{figure}

\autoref{fig:design:system_arch} shows the architecture of the Spoki probing
system. Packets from the \textit{telescope} (left side of the figure) are captured by a \textit{router} component using libtrace~\cite{libtrace}, which can ingest packets in parallel threads. It extracts packet information such as the flow tuple and payload, and routes packets to a pool of \textit{shards}. Routing decisions are based on consistent hashing~\cite{dhjkl-dahak-07, kllpl-chrtd-97} to ensure that packets with the same IP address are routed to the same shard. This approach allows us to implement rate limiting of packets that are sent to a specific target and potentially cache information for the duration of a handshake or across multiple packets.

Each shard decides which targets to probe and forwards the data to the
\textit{broker}. A broker connects to one or more Scamper daemons that handle
construction and sending of message probes. Scamper can be easily managed via a
socket connection, and it supports various probing methods, including our newly implemented \textit{synack} probe, which we use to accept TCP connections. The \textit{decoder} parses probe confirmations sent by Scamper and writes them to disk.
In addition to probe logs, Spoki writes all data from each packet including the payloads and an id that links packets to their respective~probes.

To analyze two-phase scanners, Spoki implements the decision tree in \autoref{fig:scan:taxonomy}. 
It uses simple rules to decide how to reply: \one TCP SYN messages are answered with a TCP SYN ACK message, \two ACK messages are answered with a RST message, and \three all other TCP packets are ignored.
Note that the SYN-ACK responses sent out by Spoki are irregular themselves as they do not carry TCP options. %

Probes are orchestrated by Spoki and sent asynchronously by Scamper. We avoid
sending multiple packets to the same target simultaneously by performing rate
limiting: 
we do not answer SYNs from the same source address until Scamper confirms
that it sent out the reply. 
Since care needs to be taken not to miss
establishing connections for the second (regular) phase after having replied to
an irregular SYN, we rate limit the replies to regular and irregular TCP SYNs
separately.

Rate limiting means that Spoki likely does not reply to all packets from scanners that either traverse our IP space horizontally or the ports of a single host vertically without using a permutation similar to ZMap.
For horizontal scans this is not an issue, as scanners are likely to send the same payload to all hosts. Unanswered SYNs could still be matched to the same scan process.
In contrast, Spoki might miss payloads in vertical scans, which could be avoided by adding the destination port as an additional factor to decide when to respond.

Phases are matched via time correlations between the first and second phase, \ie the second phase must start after we replied to an irregular packet in the first phase.
In addition, we only consider related only packets that arrive within 10 minutes of each other.
While both phases can originate from different ports we assume they are sent by the same source address and target the same host and port in our network.
Again, Spoki sends responses fast for two reasons.
First, to overcome TCP~timeouts:
In preliminary tests, we found that we need to respond in less than five~seconds. 
Otherwise, we will lose $\approx$ 45\% of complete handshakes because scanners will timeout and send RSTs instead of ACKs.
5~seconds for any of the millions of requests per minute is challenging.
Second, to prevent informed scanners to detect abnormal activities: Given our
attacker model where two-phase scans are largely used by malicious actors, we
may assume advanced attackers that try to detect monitoring systems based on delay-based heuristics.

\subsection{Evaluating the Scalability of Spoki}
\label{sub:scalability}

Spoki is built around four main tasks: \one packet \textit{ingestion} via the router, \two  \textit{core} evaluation performed by the shards, \three \textit{logging} events to disk via the collector, and \four \textit{probing} handled by the broker via Scamper.
We evaluate the performance of Spoki using a Linux server (Ubuntu 20.04.1 LTS), which features two AMD EPYC 7702 64-Core processors and 512GB of RAM. A local pair of virtual Ethernet interfaces route packets between the packet source and Spoki. We use ZMap to generate TCP SYN packets, passing the MAC address of the destination interface along with a probing rate and a range of IP addresses to use as a source. 
We use two /9 prefixes, one source and one destination prefix. 
We will publish detailed configurations along with Spoki's source code.

\begin{figure}
  \centering
  \includegraphics[width=\columnwidth]{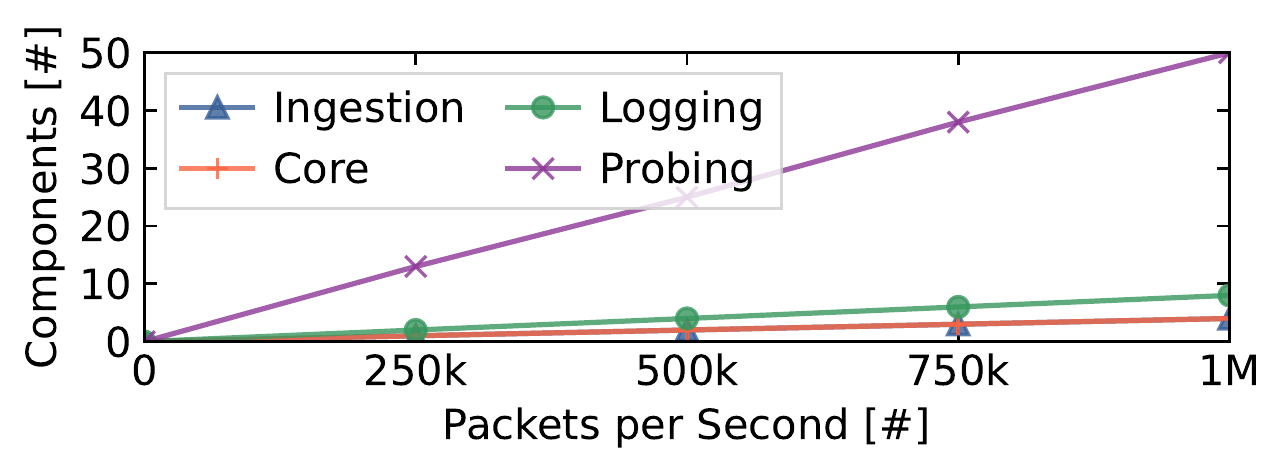}
	\caption{The scalability of Spoki components.}\vspace{-4pt}
  \label{fig:design:system_eval}
\end{figure}

The highest packet rate we tested is 1Mpps. This value is derived from the traffic volumes we observe in the UCSD network telescope, which receives around 800kpps TCP at the beginning of 2021. While the traffic volume will continue to rise, our goal is to show that Spoki can meet the challenge--and scale to larger volumes when needed. The results are shown in \autoref{fig:design:system_eval} with the number of packets per second on the x-axis and the required components (\ie one thread and one Scamper instance) to handle the load on the y-axis.

\paragraph{Ingestion} In this setup, Spoki only deploys the packet ingestion based on libtrace alongside actors that count the received packets. 
One Spoki component is able to handle about \num{250000} packets per second.
This scales linearly, and 1Mpps can be handled by only four components.

\paragraph{Core} Here, we add one shard per ingest thread. The probe requests generated by shards are collected by actors that count them. The packet processing rate matches the ingestion rate.

\paragraph{Logging} I/O puts additional strain on the shards and doubles their outgoing message volume, one message to the probers and one message to the logger. As a result, a single shard can no longer handle \num{250000} messages per second. Our evaluation shows a doubling for the number of shards: Scaling up to 1 million messages per second needs eight shards.

\paragraph{Probing} Spoki parallelizes the use of multiple Scamper instances by deploying multiple networking loops, one for each instance. To reduce the resources needed for the communication with Scamper these loops can use unix-domain sockets for communication. This way, we can scale easily with about \num{20000} probes per second per instance. To reach the targeted probing rate we deploy 50 Scamper instances, which can be optimized in the future.

\paragraph{Response delay} 
We target a packet rate of \num{100000} probes per second and collect about 6 million data points between packet receipt and the response probe. The mean and 99\% confidence intervals are 0.08 milliseconds. This underscores that Spoki is fast and does not miss accepting a handshake.

\subsection{Deployment at a Reactive Telescope}
\label{sub:consequences}

We deploy Spoki in four \texttt{/24} IP prefixes across the US and Europe on commodity hardware.
The machine in the US has 8 cores available and 16GB of RAM, running a single instance with eight shards.
The CPU usage is between 5\% and 15\% per core and the processes have a resident set size (RSS) of $\approx$80k bytes memory (about 2GB of virtual memory). 
Three Spoki instances run in the EU, each with its own /24 prefix, on a single machine with 16~cores and 31GB RAM.
The overall CPU usage is 10\% to 20\%, and the processes use between 50k and 70k RSS (slightly above 2GB virtual memory).
 Scamper sends probes from a separate machine (US), or the same host (EU), which performs both listening and probing. We separately collect PCAP files of the live data.

It is an integral part of Spoki to send packets with source addresses from the IP space of the telescope. Since network telescopes do not bind specific sockets, they cannot send packets themselves. In practice this
requires issuing spoofed packets from a  host that is connected to the same upstream network to remain indistinguishable to externals.
Since reactivity might have unexpected consequences and cause traffic to change for the active versus the passive telescope IP space, we deploy Spoki  in this study only on a small portion of each telescope. In the future, this will allow us to analyze potential divergence
between its passive and reactive components.

Spoki causes at least two extra packets for each two-phase scanner it answers. In some scenarios, this number is much higher. Some scanners answer the SYN-ACK reply with a RST as part of the first phase. Moreover, the ACK packet sent during the second phase may be retransmitted.

Overall, running Spoki is closer to a telescope than a honeypot, since its responses are minimal and do not include any payload. 
Our vantage points were previously dark and do not share a prefix with a larger active network; 
as such, they do not have a bias towards specific traffic.

\section{Characterizing Two-Phase Scanners}
\label{sec:eval}

We now want to analyze two-phase scanners and understand how they operate and  how they differ from regular one-phase scanners. Our main observation points are two reactive telescopes operating in Europe and the U.S. from Apr. through Jun. 2020. We complement our measurements using flow data from the European IXP and the Asian ISP to strengthen selected analyses, as well as from two additional telescopes for control purposes. 

In our analysis we exclude traffic from well-known scanning projects (BinaryEdge, Censys, Kudelski, Rapid7, security.ipip, Shadowserver, and Shodan) as well as a few other IP addresses attributable to research measurements  (e.g., via DNS names). This sanitization filters out 1.21\% of the two-phase events, 3.44\% of events that start with an irregular SYN, and 8.43\% of events that start with a regular SYN.

\subsection{Scanning Patterns}
\label{sec:eval:behavior}

\begin{figure}%
    \centering
    \includegraphics[width=\columnwidth]{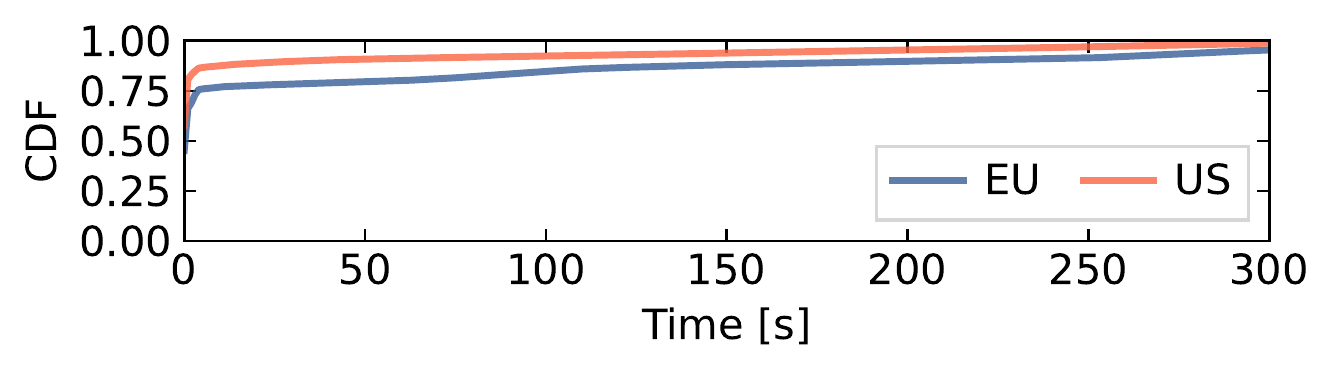}
	\caption{Temporal distribution of return for two-phase scanners: 75\% of second phases are initiated within a few seconds.\vspace{-8pt}}
    \label{fig:eval:returntime}
\end{figure}

Two-phase scanners are characterized by returning with a regular TCP SYN after a successful irregular scan in the first phase. 
\autoref{fig:eval:returntime} shows the CDF for the time lag between our SYN ACK (first phase) and the regular SYN (start of second phase).
Such scanner reactions happen fast: 75\% of the scanners return in less than a second (US) or three (EU). Our event detection considers return times of up to ten minutes and misses virtually none.

Inspecting the tails of slowly returning scanners in more depth reveals two types of behavior. A larger number fluctuates in return time  up to 100 seconds, which may be  due to scheduling under the system load of many replies from the first scanning phase. A smaller group admits  constant, significant delays. We attribute this to implementation or configuration details of the scanning software. 
No correlations with source locality or targets (IP, port) were visible. 

\begin{figure*}%
  \begin{subfigure}{.25\textwidth}
    \centering
    \includegraphics[width=\linewidth]{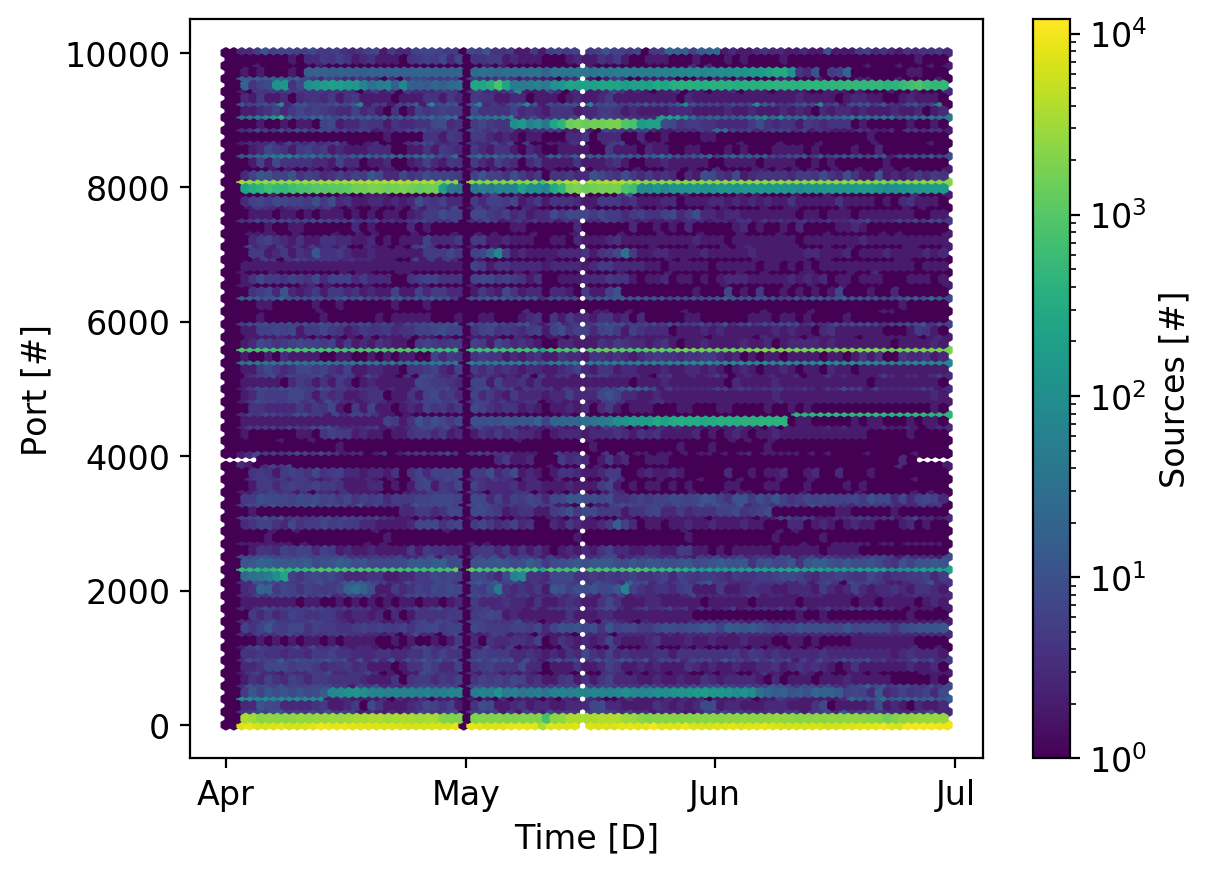}
    \caption{Two-phase (US).}
    \label{fig:eval:goals:ports:sources:us:two}
  \end{subfigure}\hfill
  \begin{subfigure}{.25\textwidth}
    \centering
    \includegraphics[width=\linewidth]{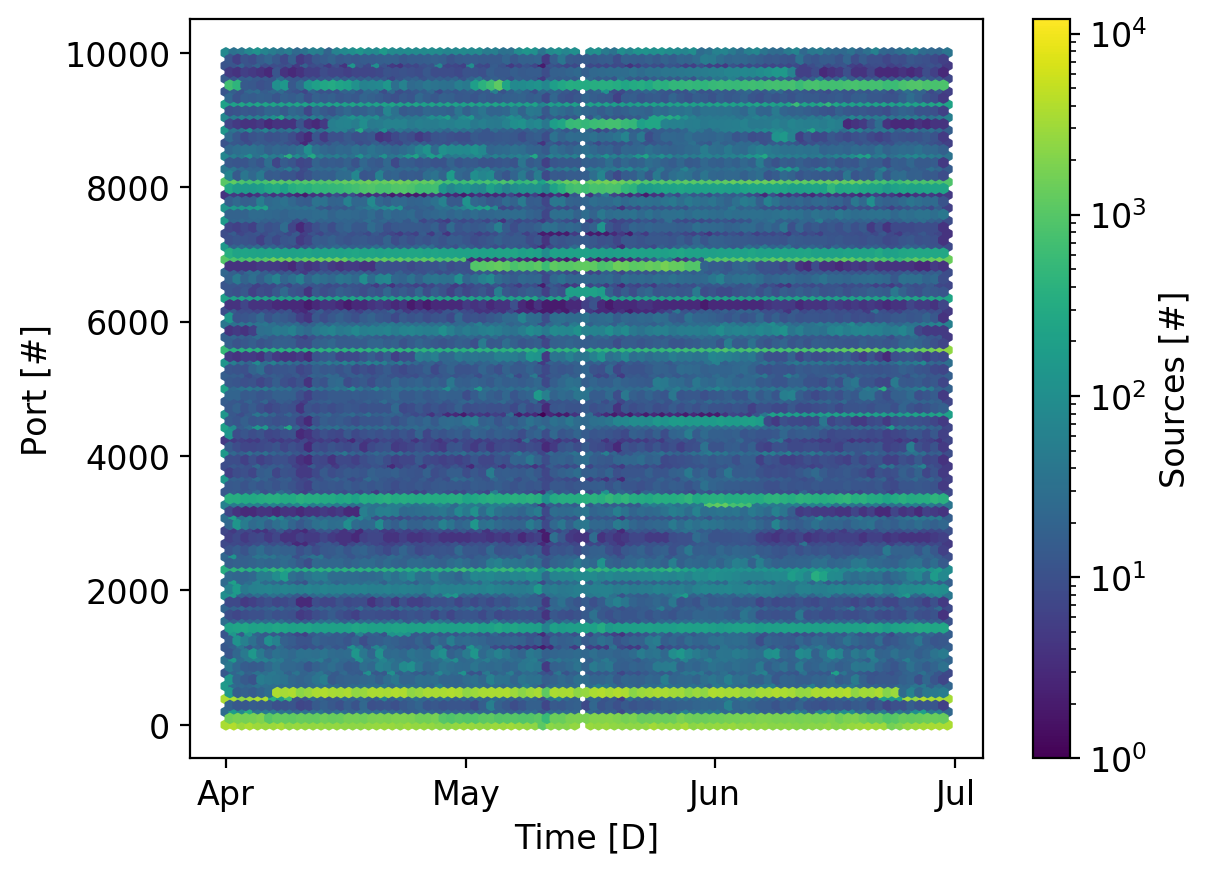}
    \caption{One-phase (US).}
    \label{fig:eval:goals:ports:sources:us:one}
  \end{subfigure}\hfill
  \begin{subfigure}{.25\textwidth}
    \centering
    \includegraphics[width=\linewidth]{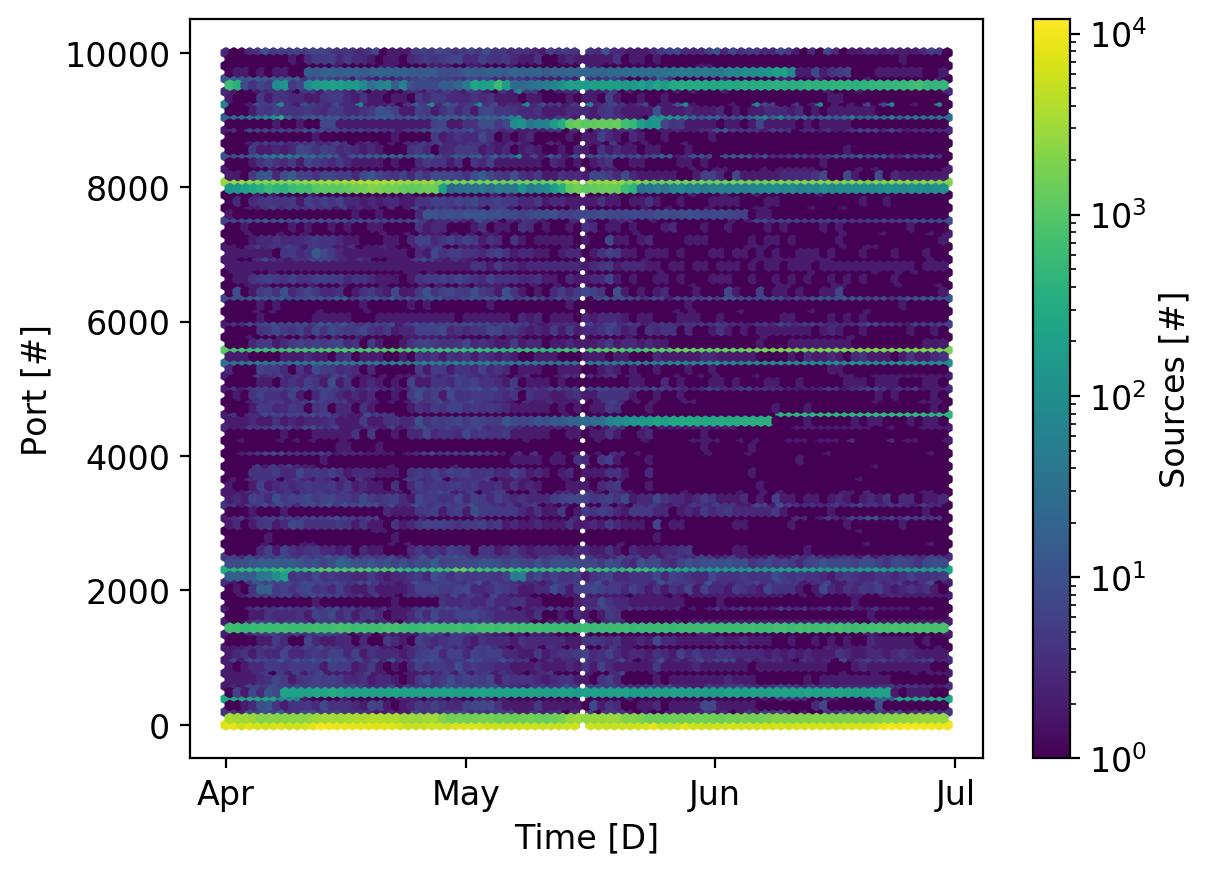}
    \caption{Two-phase (EU).}
    \label{fig:eval:goals:ports:sources:eu:two}
  \end{subfigure}\hfill
    \begin{subfigure}{.25\textwidth}
    \centering
    \includegraphics[width=\linewidth]{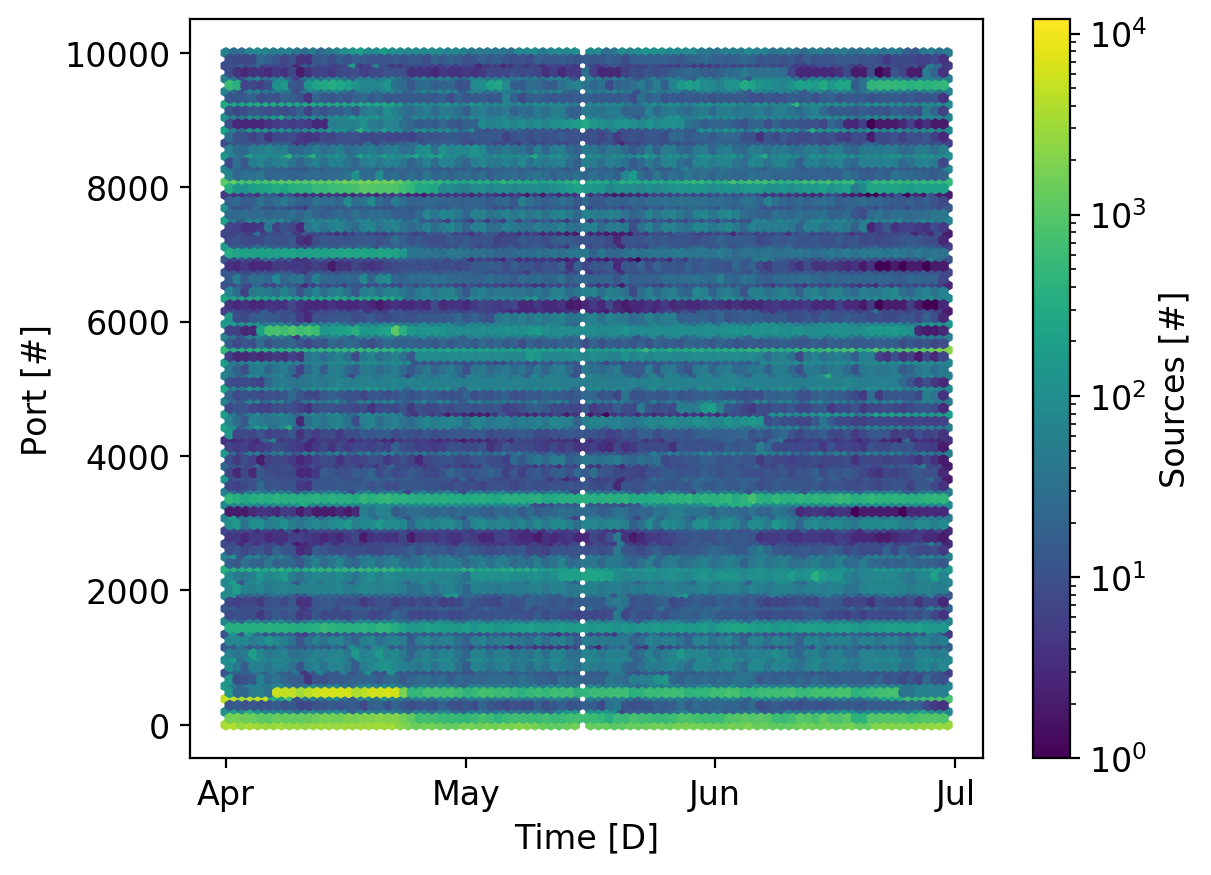}
    \caption{One-phase (EU).}
    \label{fig:eval:goals:ports:sources:eu:one}
  \end{subfigure}\hfill
  \caption{Unique sources per day that probed the first 10k ports. Two-phase scanners are almost invisible beyond selected ports.}
  \label{fig:eval:goals:ports:sources}
\end{figure*}

\begin{figure*}%
  \begin{subfigure}{.25\textwidth}
    \centering
    \includegraphics[width=\linewidth]{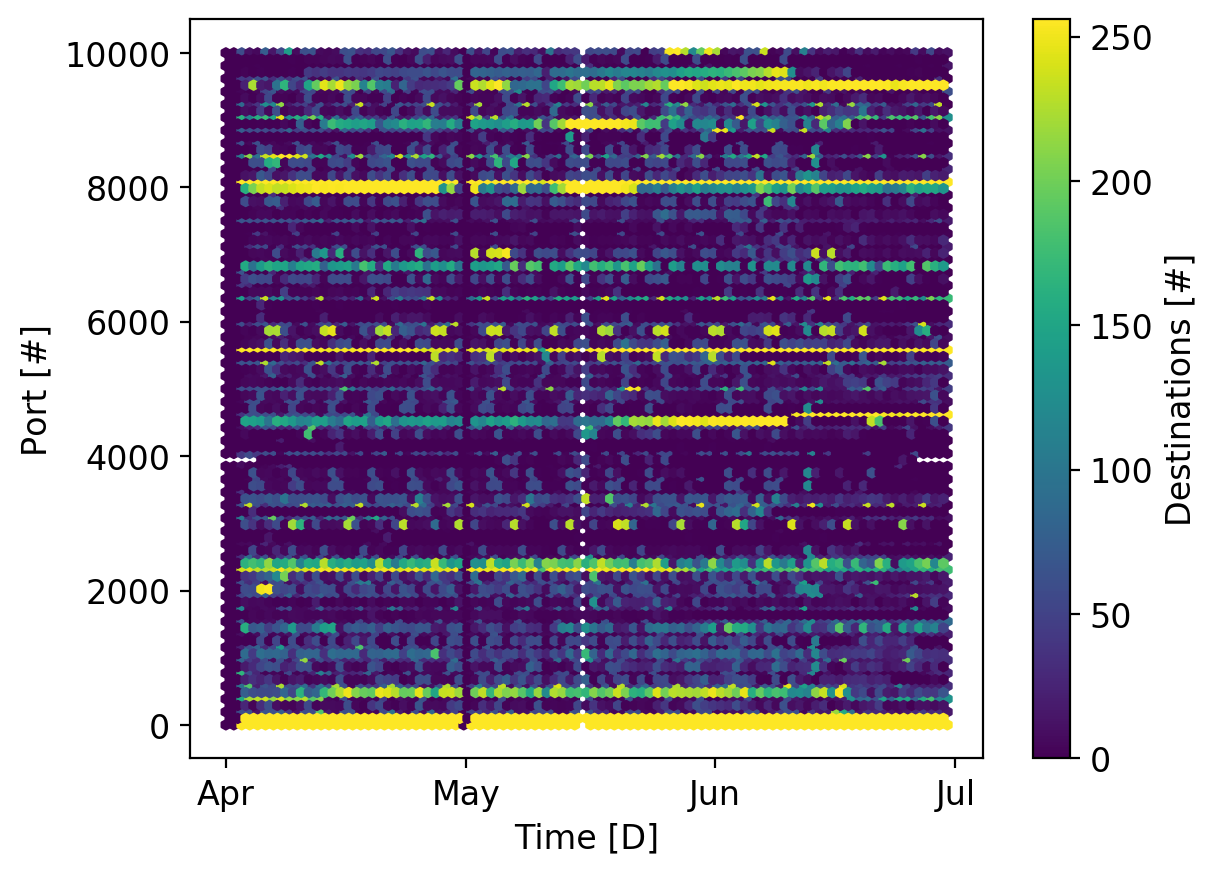}
    \caption{Two-phase (US).}
    \label{fig:eval:goals:ports:target:us:two}
  \end{subfigure}\hfill
  \begin{subfigure}{.25\textwidth}
    \centering
    \includegraphics[width=\linewidth]{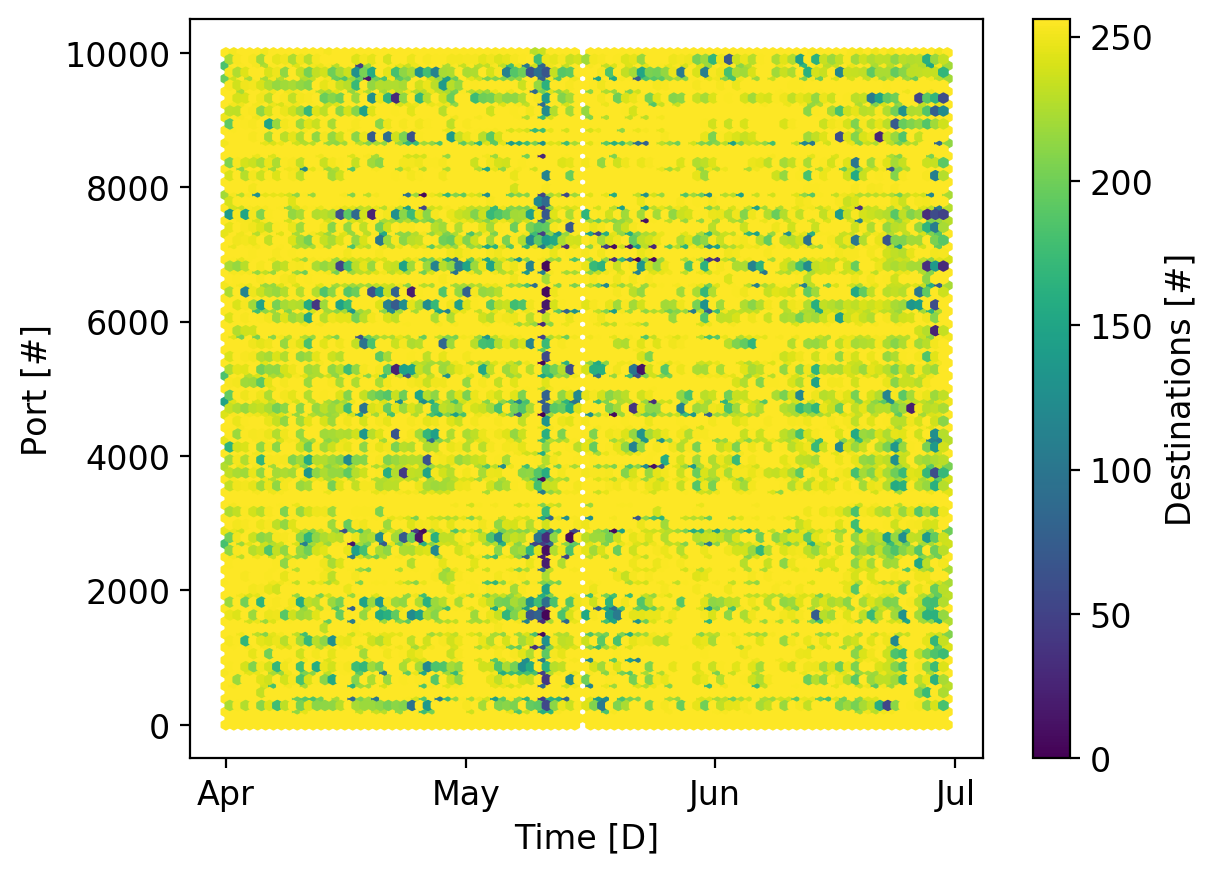}
    \caption{One-phase (US).}
    \label{fig:eval:goals:ports:target:us:one}
  \end{subfigure}\hfill
  \begin{subfigure}{.25\textwidth}
    \centering
    \includegraphics[width=\linewidth]{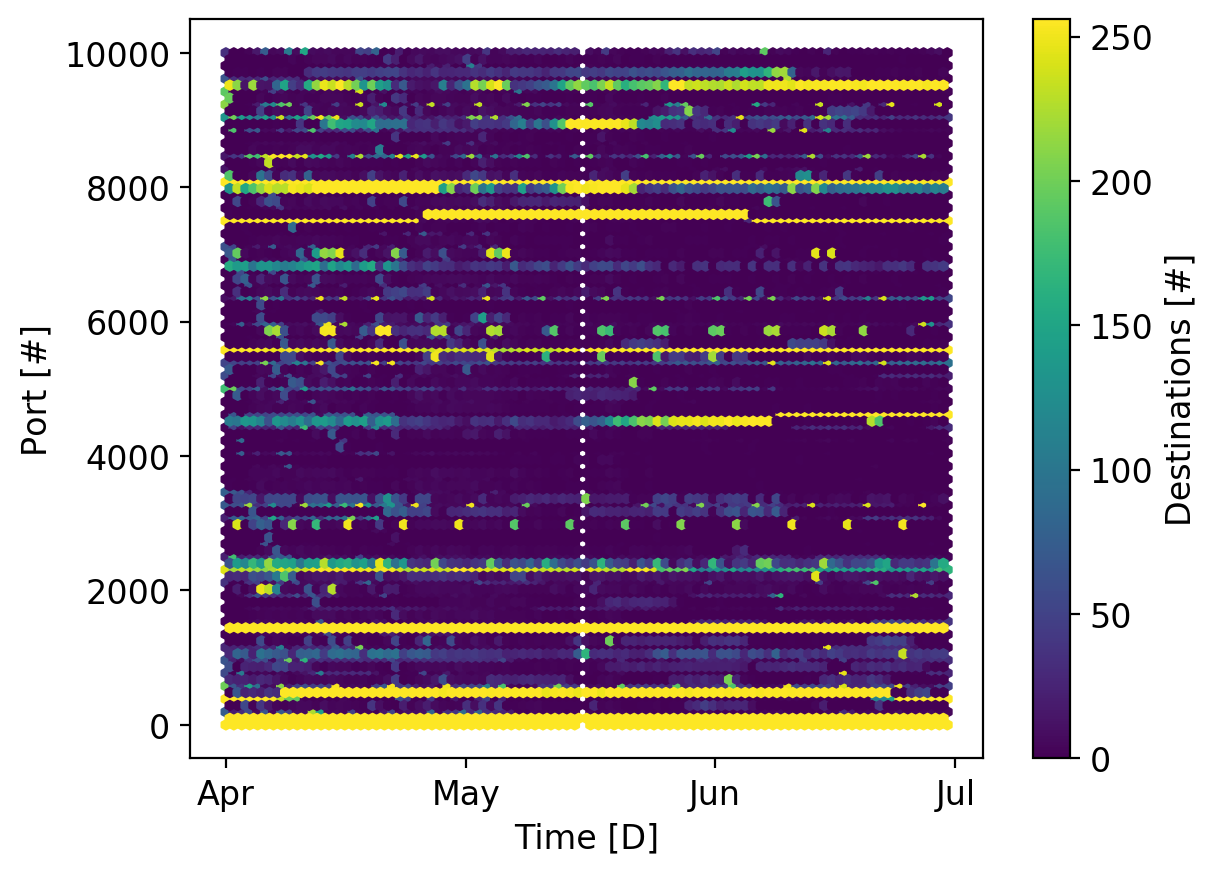}
    \caption{Two-phase (EU).}
    \label{fig:eval:goals:ports:target:eu:two}
  \end{subfigure}\hfill
    \begin{subfigure}{.25\textwidth}
    \centering
    \includegraphics[width=\linewidth]{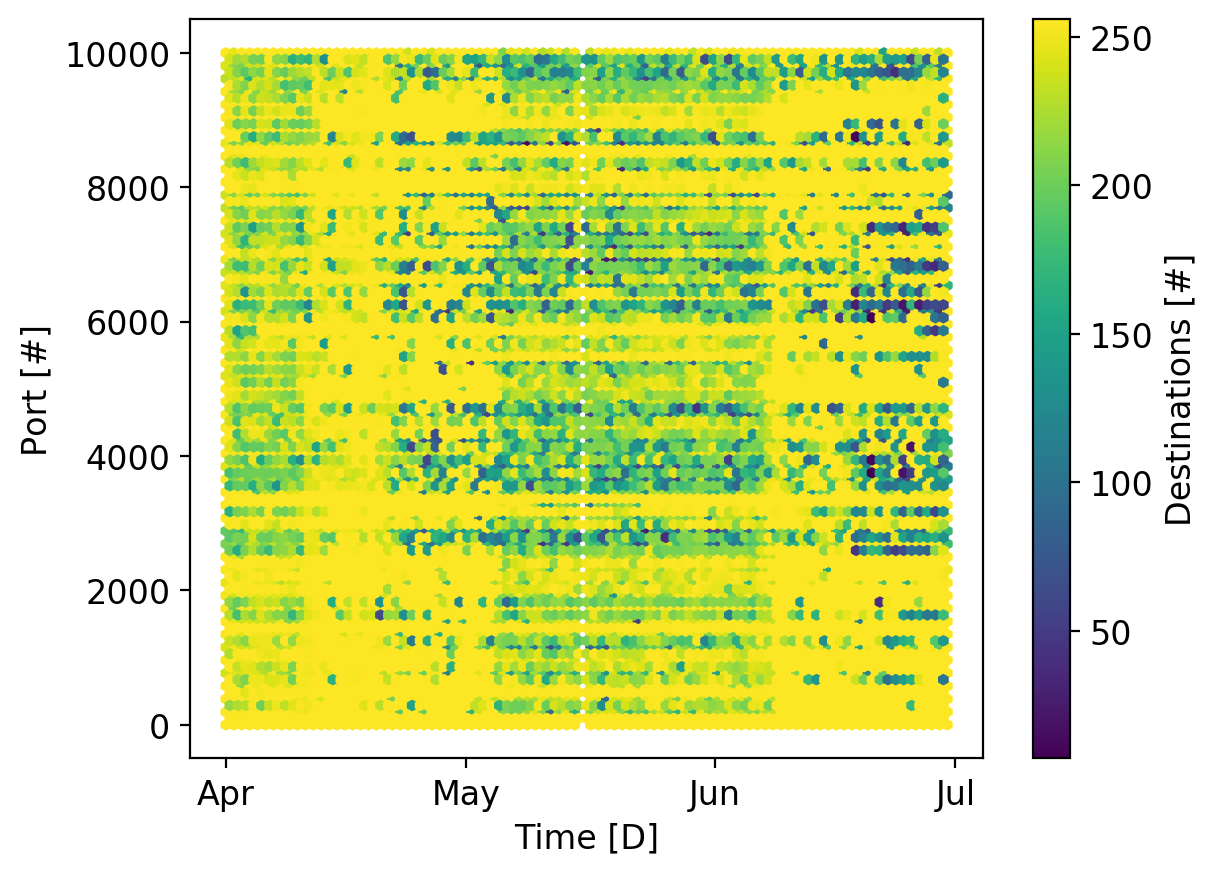}
    \caption{One-phase (EU).}
    \label{fig:eval:goals:ports:target:eu:one}
  \end{subfigure}\hfill
  \caption{Unique destinations targeted in our /24 subnets on the first 10k ports. Two-phase scanners tend to scan all addresses of the subnets on selected ports and in specific time frames. Their service targets do not appear globally uniform.}
  \label{fig:eval:goals:ports:target}
\end{figure*}

Next we compare the activities of two-phase and regular one-phase scanners. We want to understand whether two-phase scanners simply use a different tool-set but follow the same goals as regular scanners, or whether they have specific targets and distinct behavior.    
We first examine the scanning flows towards specific services with two metrics: how many sources are interested in a specific port on a given day, and how many of our hosts were scanned on a given port and day.

\autoref{fig:eval:goals:ports:sources} visualizes the two-phase scanning activities for the lowest 10,000 ports and compares them with regular single phase scans. In all cases, we see the same peak intensity of about 10,000 different sources per day scanning specific ports. The signatures of one- and two-phase scanners, however, largely differ. While one-phase scanning activities cover wide port ranges often at intermediate intensity, the two-phase scanners appear very focused in time and on specific ports. 

This initial observation aligns with the scan coverages in our telescopes. \autoref{fig:eval:goals:ports:target} shows the number of scanned hosts in our networks per port and day. Clear scanning patterns become visible for the two-phase scanners, whereas the one-phase scanning remains diffuse.
This targeted behavior shows intent among the scanners as well as a consensus on what to probe.
The next section further examines why two-phase scanners contact us.

Our analysis shows that two-phase and one-phase scanners both perform vertical scans. However, two-phase scanners focus more significantly on horizontal scans and engage much less in vertical scans.
While previous work noticed a growing focus on vertical scans it either did not differentiate the behavior based on the scanner types~\cite{wktpl-hs-18, apt-bhs-07} or did not examine scanning in this context~\cite{aabbb-umb-17}.

\begin{figure}[!b]
    \centering
    \includegraphics[width=\columnwidth]{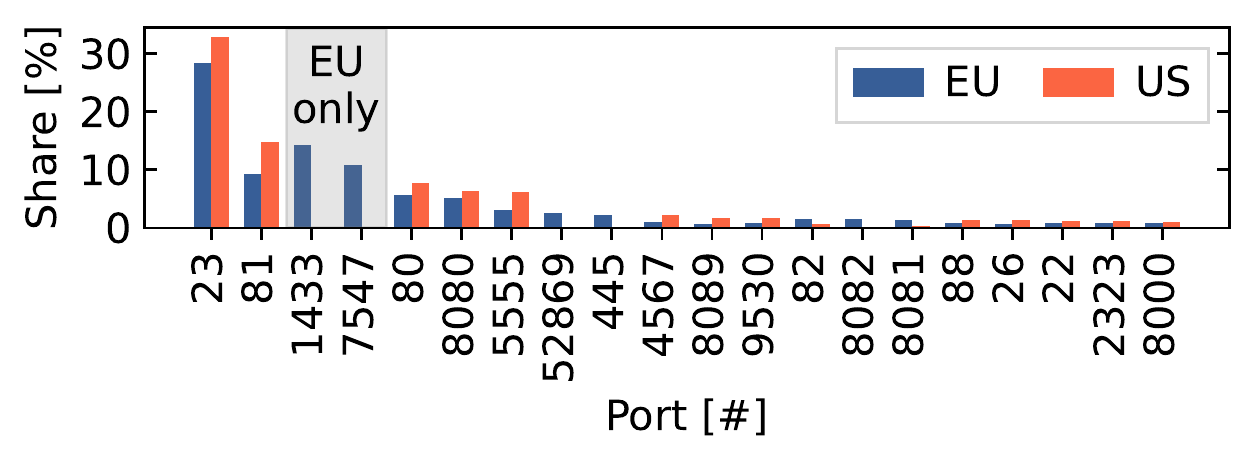}\vspace{-4pt}
	\caption{Distribution of top-20 ports. 20\% of the scans are only visible at the EU vantage point.}
    \label{fig:eval:portstat}
\end{figure}

The distribution of the topmost visited service ports is displayed in \autoref{fig:eval:portstat}. We find about 30\% of all two-phase scans target the {\tt telnet} port 23, which was also the first contact point for Mirai. This is about half of the share observed by Heo et al.~\cite{wktpl-hs-18} in 2018. We only see around 1\% companion scans on port 2323, which was a unique signature of Mirai~\cite{aabbb-umb-17}. For both ports we observe frequent (up to 200) connection retries in the second phase of the scan, which Spoki answers with a TCP reset. These retries could be attributed to the Mirai strategy of probing for multiple credentials.   
On the overall it appears as if Mirai  persists in the scanning ecosystem, but slowly gets replaced by scanner variants of altered behavior.

We find that the port distribution (\autoref{fig:eval:portstat}) is highly biased with respect to the telescope location. Uniquely in Europe, $\approx$15\% of the scan events address port 1433, which can be linked to a privilege escalation exploit in SIMATIC controllers---a popular automation systems by Siemens AG, which is a dominant supplier in Europe. Originally in the old MS-SQL Spida worm, the exploit persisted as part of the SIMATIC product until a few years ago.   
10\% of the scan events in Europe link to the TR-069 exploit on port 7547, which concerns home routers mainly popular in Europe. We can  confirm this observation by inspecting irregular SYN traffic at the European IXP and Asian ISP. At the IXP, the port 7547 is the 7th most common port but only 145th in Asia. We observe irregular SYNs for port 1433 in European and Asian traffic, which may relate to successful SIMATIC deployment (\eg in India). In contrast, about 3\% of the scans at the US telescope target port 4567, which grants access to a web server for management on Verizon Actiontec home routers exploitable by a Trojan.

Analyzing the evolution of the top-5 service scans over the last 16~years, we see a shift away from traditional service ports such as the Microsoft Directory Service/SAMBA port 445 among two-phase scanners in favor of telnet as diagnosed from an Asian perspective by Heo \etal~\cite{wktpl-hs-18}.
Traditional one-phase scanners continue to concentrate on port 445 (50-60\%), followed by a wide port distribution that starts one order of magnitude lower.
Details are summarized in \autoref{apx:evolution-scan-ports}.

\subsection{Types of Interaction}
\label{sec:eval:contact}

Two-phase scanners may contact our vantage points with different intents. Some may open connections to see service announcements, also known as banner grabbing, while others deliver payloads. These payloads can be service-specific probes to elicit responses or collect information. These probes could further determine whether the service behind a port is the expected one, which is often not true~\cite{itd-liuis-21}. A third kind of scanners delivers payloads that attack the system directly.

\setlength{\tabcolsep}{4pt}
\begin{table}
  \caption{Contact types of two-phase events; shares taken relative to the contexts ``Two-phase'' or  ``w/ Payload''.}
  \label{tab:eval:contact:types}
  \centering
  \begin{tabular}{l r r r r}
    \toprule
    Event Type    & \multicolumn{2}{c}{EU} & \multicolumn{2}{c}{US} \\
    \midrule
    Two-phase     & 8,698,497 &    100.0\% & 7,899,333 &    100.0\% \\
    ~~w/o ACK     & 2,894,987 &     33.3\% & 3,665,114 &     46.4\% \\
    ~~w/o Payload & 2,126,900 &     24.5\% & 1,767,249 &     22.4\% \\
    ~~w/~~ Payload   & 3,676,610 &     42.3\% & 2,466,970 &     31.2\% \\
    \cmidrule(lr){1-5}
   \multicolumn{2}{l}{~~Details w/ Payload}
   \\
    ~~~~ASCII     & 2,155,751 &     58.6\% & 1,984,444 &     80.4\% \\
    ~~~~HEX       & 1,478,556 &     40.2\% &   339,217 &     13.8\% \\
    ~~~~Downloader&    42,303 &      1.2\% &   143,309 &      5.8\% \\
    \bottomrule
  \end{tabular}
\end{table}

\autoref{tab:eval:contact:types} lists the numbers of received two-phase events alongside the share of two-phase events that receive an ACK or a payload in the second phase. Payloads are further separated by type: \textit{ASCII}-decodable, binary data we store as \textit{HEX}, and ASCII payloads that contain \textit{downloaders}. A downloader carries code in its payload that a vulnerable service runs upon reception to download and install remote software, see \autoref{sec:payload:malware} for details. On average, payloads have been retransmitted 2.5 times in the EU and 3.2 times in the US.

The share of payloads without an ACK stands out. There is no obvious reason to return for a second phase  without completing the TCP handshake, as no relevant information is learned over the stateless scan. A small part of these events can be explained by packet loss, since we do not retransmit packets. The port distribution in this category is rather wide and resembles the initial contacts, which may be an indication of overload at the scanner side.

Servers initiate conversations for some protocols, \ie  after the TCP connection is established, they are the first to send payload. These payloads can be collected by scanners for additional information. In particular, the scanning can  wait for this payload before interacting. As an example, the telnet login starts with an authentication request from the server upon which a client can provide credentials. Roughly 70\% of the scanners in both regions contact port 23 without payload, which is consistent with banner grabbing via the telnet protocol.

Among contacts with payloads, the ASCII-decodable payloads dominate.  These scanners favor the HTTP ports 80, 8080, and 81 in both regions. The EU additionally sees a larger share of ASCII contacts on port 7547, which fits the TR-069 plain text protocol (see \autoref{sec:eval:behavior}).

Downloaders detectable in ASCII payloads are more common in the US. The most common ports include HTTP-type ports (80, 8089, 8080) as well as port 60001, which is used as a web server on some IoT devices. Consequently, these payloads are predominantly POST or GET requests.

For HEX contacts, ports 1433 (MS SQL, \textit{cf.} \autoref{sec:eval:behavior}) and 445 (active directory / SMB) dominate in the EU while the US sees contacts on ports 5555 (ADB) and 443 (HTTPS). This is consistent with the location-dependent targeting (\cf \autoref{sec:phenomenons}).

\subsection{Scan Origins}
\label{sec:eval:origin}

Two-phase scanning is globally observable in traffic flows at IXPs and ISPs as well as in network telescopes, see \autoref{sec:scanners}.
To examine the origins of two-phase scanners, we map their IP~source addresses to countries based on the NetAcuity Edge database~\cite{netacuity}.
The share of two-phase events by country for each of our vantage points are presented in \autoref{tab:eval:goals:meta:country:top_five}, for \mbox{top-5} contributors.
While both vantage points observe similar origin countries, the intensity of scans per country differs significantly. 
Compared to \cite{dbh-ivis-14}, we observe a more equal distribution across nations. China is not the major source of two-phase scans.
Most notably, Ukraine and Poland hold a significant share in the EU, while Taiwan is a major source of two-phase scans in the~US. 

We also examine the monthly evolution of traffic shares for Apr., May, and Jun. In Apr., at our US vantage point we see a country distribution similar to the averages in \autoref{tab:eval:goals:meta:country:top_five}. 
In May, Taiwan hosts the majority of scanners (14.5\%) and thus more than the US (9.5\%); this ranking swaps in Jun. again. China (6\% to 7\%), Brazil (4.9\% to 6.3\%), and Romania (4.6\% to 8.5\%) slightly increase in share.
Taking a closer look at the rise in traffic in Taiwan during May, we observe numerous addresses that each contribute few events (less than 1\%).
The majority (85\%) sits in AS~3462.
This is accompanied by a stronger interest in HTTP-like ports (80, 8080, 88, 8000, etc.) and the most common payload is a simple \texttt{GET} request (roughly 50\%). %

\setlength{\tabcolsep}{3pt}
\renewcommand{\arraystretch}{0.8}
\begin{table*}[t]
  \caption{Top-5 share of two-phase events \one ranked by country of source IP addresses and \two by source AS annotated with the AS type. The EU sees larger shares from the top IP~sources. Most sources are located in transit networks.}
  \label{tab:eval:goals:meta:country:top_five}
  \centering
  \begin{tabular}{l@{\ \ } l@{\ } r l@{\ } r l@{\ } r l@{\ \ } r}
    \toprule
    & \multicolumn{4}{c}{Ranked by source IP address} & \multicolumn{4}{c}{Ranked by source AS} \\
    \cmidrule(lr){2-5}
    \cmidrule(lr){6-9}
    & \multicolumn{2}{c}{EU} & \multicolumn{2}{c}{US} & \multicolumn{2}{c}{EU} & \multicolumn{2}{c}{US} \\
    \midrule
    1. & Ukraine (UA)       & 13.37\% & United States (US) & 11.93\% 
    & Transit/Access (UA) & 10.11\% & Transit/Access (CN) & 9.36\% 
    \\
    2. & Russia (RU)        & 11.54\% & Taiwan (TW)        & 11.04\% 
    & Enterprise (PL)     &  7.48\% & Transit/Access (UK) & 5.60\% 
    \\
    3. & China (CN)         & 11.46\% & Romania (RO)       &  6.62\% 
    & Enterprise (RU)     &  7.29\% & Transit/Access (KR) & 3.94\% 
    \\
    4. & Poland (PL)        & 11.22\% & China (CN)         &  6.53\% 
    & Transit/Access (TW) &  5.19\% & Hosting (US)        & 3.19\% 
    \\
    5. & United States (US) &  7.32\% & Russia (RU)        &  5.56\% 
    & Transit/Access (CN) &  4.01\% & Transit/Access (CN) & 2.91\% 
    \\ 
    \bottomrule
  \end{tabular}
\end{table*}

The EU telescope observed even more variation. In Apr., the biggest share is held by Ukraine~(17.8\%), followed by China (13\%) and the US (8.3\%). Traffic from Russia (5\% to 15.6\%) and Poland (3.5\% to 17.4\%) increases significantly over the months. Similar to events observed at the US telescope, the share of events originating from Taiwan peaks in May (8.6\%). Overall, two-phase scanning activities appear rather volatile and unbound to specific infrastructure. 

To analyze the type of networks that host two-phase scanners, we map each source IP address to its origin autonomous system (AS).
\autoref{tab:eval:goals:meta:country:top_five} lists types and countries of the top five ASes observed at each vantage point. The AS types are based on the CAIDA AS Classification data set~\cite{caida-relationships}.
The top contributing ASes are transit and access providers (70\%). At our EU vantage point, we observe two enterprise networks located in Poland and Russia, while we only see a single hosting provider (Digital Ocean) among the top 5 at our US telescope.

We examine the source addresses hosted in transit/access ASes in  detail to understand whether they belong to residential end hosts instead of server-based infrastructure.
Using reverse DNS queries we find that most of the names match the patterns of access nodes (\eg an IP address followed by a domain).
In most of these ASes, a large number of IP addresses (10k+) send out few probes each.
Two~ASes, however, differ: AS~48081 and AS~9009 observed in the EU and US, respectively, in which we only see a few source IP addresses that massively send out two-phase scan packets.
We now inspect these further.

All traffic originated from the Ukrainian AS~48081 relates to a single scanner. 
Using HTTP banner grabbing we find that the scanner runs on a Mikrotik~DSL router, which is known to be vulnerable to a number of exploits~\cite{jspa-mdlrh-20}.
Most of the probing packets target the highly vulnerable services TR-069 (7547) and Telnet (23).
Potential ports of HTTP servers (\ie 80, 8080, 8002, 8001, 81, 82) are scanned one order of magnitude less. 

AS~9009 is located in Romania and hosts nearly 80~scanners. Up to five of the scanners contribute most of the traffic. These heavy hitters exhibit sequential IP~addresses and generic residential router names.
The distribution of targeted ports suggests systematic scanning. In particular, two addresses scan a large variety of ports with similar frequencies. These scanners exclude ports 23 and 7547.

One option to host scanners is the cloud. We examined the top ten observed ASes of hosting providers at both vantage points. The largest share is held by the hosting provider represented in the top five in the US. Only two more hosting providers held more than 1\%. From this we conclude that a two-phase scanner infrastructure in the cloud is not popular in practice, similar to the observation of Heo \etal~\cite{wktpl-hs-18} for telnet scans. 
The efficiency of stateless scans combined with exploitable victims eases infrastructureless attacks.
Attackers hence tend to infiltrate end hosts and create a (P2P) botnet.

\begin{figure}[b]
  \centering
  \includegraphics[width=\linewidth]{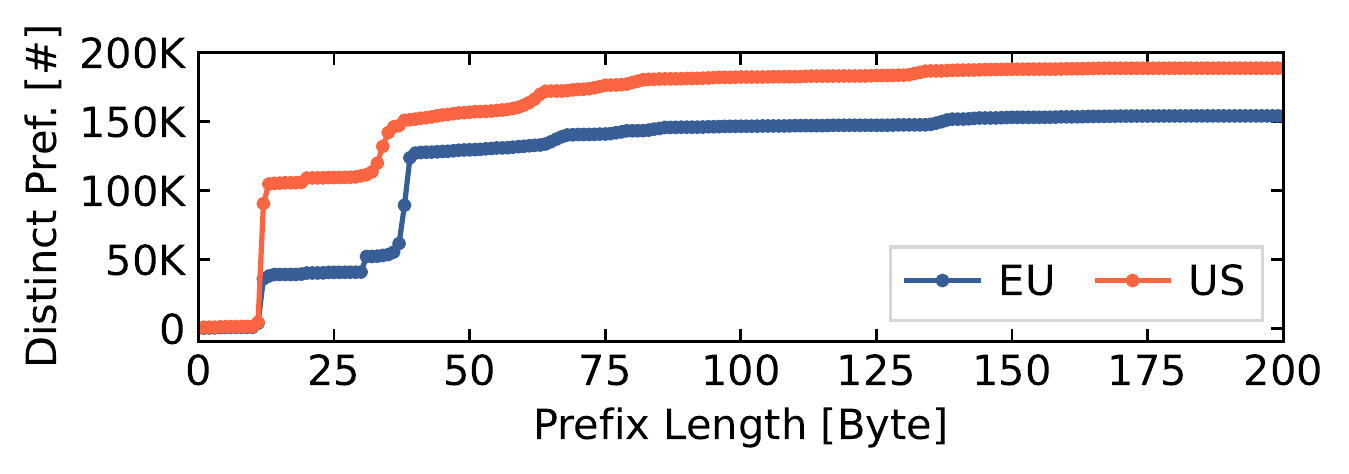}
  \caption{Distribution of distinct payload prefixes: Steps around 12 bytes and 30 bytes hint at common protocols.}
  \label{fig:eval:goals:payloads:distict-prefixes-200}
\end{figure}

\section{Transport Payloads}
\label{sec:payload}

Spoki collects payloads from TCP ACK packets that complete a handshake, which is one of its unique features compared to a traditional network telescope.
\autoref{tab:eval:contact:types} lists the share of events that contacted us with a specific payload type.
We observe \num{166035} and \num{190905} unique payloads in the EU and the US respectively, where 38\% and 41\% are ASCII-decodable.
The most frequently scanned port (port 23) almost never received a payload, \cf \autoref{sec:eval:contact}.

\setlength{\tabcolsep}{3pt}
\renewcommand{\arraystretch}{0.7}
\begin{table*} %
  \caption{Observed ASCII and hex payloads ranked by maximum share.}
  \label{tab:eval:goals:payloads:prefixes:ascii}
  \centering
  \begin{tabular}{l@{\ \ } r@{\ \ } l@{\ } r@{\ \ } l@{\ \ } l@{\ \ } r@{\ \ } l@{\ \ } r@{\ \ } l@{}}
    \toprule
       \multicolumn{5}{c}{ASCII payload} & \multicolumn{5}{c}{Hex payload} \\
    \cmidrule(l){1-5} \cmidrule(l){6-10} 
       & \multicolumn{2}{c}{EU} & \multicolumn{2}{c}{US} & & \multicolumn{2}{c}{EU} & \multicolumn{2}{c}{US} \\
    \cmidrule(lr){2-3} \cmidrule(lr){4-5} 
    \cmidrule(lr){7-8} \cmidrule(lr){9-10} 
    Payload prefix                       & Share   & Ports           & Share   & Ports  
     & Payload prefix                   & Share   & Ports           & Share   & Ports
    \\
    \midrule
    \texttt{GET / HTTP/1.1}             & 73.50\% & 7547, HTTP$^1$  & 46.75\% & HTTP$^1$ 
     & TDS7$^3$ Pre-login               & 74.52\% & 1433            &  1.16\% & 1433
    \\
    \texttt{GET login.cgi}              & 14.64\% & HTTP$^1$        & 21.18\% & HTTP$^1$ 
     & TLS Client Hello                 &  4.55\% & 443, 8443       & 37.80\% & 443, 8443 
    \\
    \texttt{GET / HTTP/1.0}             &  1.10\% & HTTP$^1$        & 10.49\% & HTTP$^1$, 10255 
     & ADB$^4$ Connect                  &  4.97\% & 5555            & 37.01\% & 5555       
    \\ %
    \texttt{POST /}                     &  3.72\% & 52869, HTTP$^1$ &  4.12\% & HTTP$^1$ 
     & SMB Negotiate                    & 11.04\% & 445             & --      &            
    \\
    Debugging port$^2$                  &  1.09\% & 9530            &  2.04\% & 9530 
     & PSQL/UPnP                        &  0.35\% & 5432            &  3.10\% & 5432, 5000 
    \\
    Hardcoded auth$^2$                  &  0.95\% & 4567            &  3.13\% & 4567 
     & TSAP                             &  0.45\% & 102             &  1.42\% & 102        
    \\ %
    \texttt{SSH-2.0-}                   &  0.93\% & 22, 2222        &  1.65\% & 22, 2222 
     & MongoDB                          &  0.27\% & 27017           &  1.21\% & 27017      
    \\ %
    \texttt{GET /s} (e.g., shell)       &  0.71\% & 8983, 60001     &  2.01\% & 8983, 80, 60001 
     & \textit{Unknown}                 &  0.16\% & 28967           &  1.15\% & 28967     
    \\
    \bottomrule
    \multicolumn{5}{l}{\footnotesize $^1$HTTP-like ports such as 80, 8080, 8081.} & \multicolumn{5}{l}{\footnotesize $^3$Tabular Data Stream Protocol (TDS) used by Microsoft SQL.} \\
    \multicolumn{5}{l}{\footnotesize $^2$Strings related to exploits of embedded devices, \eg DVRs, NVRs, and~IP~cameras.} & \multicolumn{5}{l}{\footnotesize $^4$Android Debug Bridge (ADB).} \\
  \end{tabular}
\end{table*}

For a detailed analysis, we group payloads by longest common prefixes of the first $N$ bytes, since common prefixes hint at similarities such as identical protocols.
We chose a maximum prefix length of 200 bytes since both vantage points dominantly captured payloads with 200 bytes or fewer. 
\autoref{fig:eval:goals:payloads:distict-prefixes-200} depicts the number of distinct prefixes per payload length. The prefixes diverge significantly after 12~bytes (\ie \texttt{GET / HTTP/})
and after 31~bytes, exhibiting more branches in the EU data (\ie \texttt{GET / HTTP/1.1\char`\\r\char`\\nHost: IP.AD.DR:PORT} with the IP address matching the first three octets of our IP prefix). 
At about 35 bytes a few prefixes from the US branch, but are hard to pin down to a specific payload.
The large EU increase terminates at 40 bytes, which  marks a hex payload of the Tabular Data Stream Protocol.

These striking jumps suggest large shares of similarly structured payloads. To identify further commonalities, we build a prefix tree and match trivial components (\eg destination IP~addresses). Based on the common payload prefixes, combined with targeted ports, we identify the strongest contributing protocols as listed in \autoref{tab:eval:goals:payloads:prefixes:ascii}.
At both vantage points, the majority of the observed payloads are HTTP GET requests version 1.1. 
Port 7547 stands out in the EU, which is related to the TR-069 vulnerability.
The prefix \texttt{GET /s} covers \texttt{GET /shell}, including further commands that switch directories, download a shell script or a binary, and give permission to execute (example in \autoref{lst:eval:payload:get:shell}). 
\begin{lstlisting}[numbers=left,frame=single,framexleftmargin=15pt,caption=Query of a GET request observed to \texttt{/shell} on port 37215. It downloads and executes \texttt{arm7}.,label=lst:eval:payload:get:shell]
cd /tmp; rm -rf *; \\
  wget http://IPv4/arm7; \\
    chmod 777 arm7; ./arm7 rep.arm7
\end{lstlisting}

\subsection{Assessing Maliciousness}
\label{sec:payload:malicious}

We first analyze the maliciousness of payloads semi-manually by extracting payload data, decoding it, and comparing to known exploits.
While this generates a basic overview over a subset of  payloads, we follow a systematic approach thereafter using blocklists and a threat intelligence provider.

We can tell from the payloads that 150k POST events in Europe (600 in the US) relate to a UPnP exploit of Realtek that can lead to code injection.
Two other frequent payloads send hex strings related to exploits in embedded devices, see ``Debugging port'' and ``Hardcoded auth'' in \autoref{tab:eval:goals:payloads:prefixes:ascii}.

We decoded binary payloads using Wireshark. Aside from the payload sent to port 28967, which we could not decode, target services (or devices) relate to known exploits. In the EU, a dominant share of two-phase scanners is hitting port 1433 (Microsoft SQL/SIMATIC) while the share in the US is evenly distributed between 443 and 5555 (TLS and Android Debug Bridge).
Some Android devices allow attackers access on this port, \eg to install cryptocurrency miners~\cite{tbakb-iiisc-20, lls-abiar-21}.
Port 27017 (MongoDB) was targeted by hackers in 2020, who copied and deleted databases to hold the data for ransom.

$\sim$3\% of the \texttt{HTTP / 1.1} payloads include the string \texttt{zgrab} and a user agent matching the default value in the ZGrab source code, \cf \cite{zmap--github-scanner-go}. We observe two variations of the ``Host'' headers: one uses the form HOST:PORT and favors port 80, the other lists an integer and targets up to 5~ports equally.

Overall, we see payloads that are clearly malicious (\eg carrying shell code) as well as more harmless ones (\eg those using the ZGrab user agent). However, payloads that look benign at first glance could be used for reconnaissance scans by malicious actors.

For a systematic cross verification, we compare source IP addresses to blocklists and data from GreyNoise (GN)~\cite{greynoise}, a threat intelligence provider.
We acknowledge that these systems have  limitations~\cite{ldpmv-rtlca-19} but we also emphasize that there is currently no perfect solution~\cite{fvgpn-bbtdo-21}. We try to minimize systematic errors by using diverse sources as a loose form of verification.
\textit{URLhaus}~\cite{urlhaus} %
shows a source overlap of 3.0\% (7.6\% events).
MalwareWorld~\cite{malwareworld} %
includes 4.6\% of our sources (16.06\% events).
\textit{BlockIP}~\cite{pallebone} exhibits an overlap of 6.9\% (30.6\% events).
We see overlaps of 3.9\% (13.4\% events) and 2.8\% (15.6\% events) on the \textit{sshclient} and \textit{sipquery} lists collected by DataPlane~\cite{dataplane}. %
GN classifies 56\% (EU) and 70\% (US) of the two-phase payload sources as malicious. The overall share of malicious sources is comparably lower among all scanners with 38\% (EU) and 35\% (US).

In summary, all analyses of payloads and data sources reveal a notable share of malicious actors among two-phase scanners. In addition, malicious sources produce a relatively large share of the entire events.

\subsection{Analyzing Downloaders}
\label{sec:payload:malware}

We identified those ASCII-decodable payloads that perform downloads using \texttt{wget} or \texttt{curl}.
We denote the source address of the payload as the \textit{scanner} and the address targeted by the download as the \textit{hoster}.
Each event gets a \textit{name} extracted from the last path element of the URL.
The most common name we see is \texttt{arm7} with 44\% (EU) and 58\% (US), respectively.
It is followed by \texttt{mpsl} (around 10\%), \texttt{le.bot.arm7} (5-8\%), and \texttt{viktor.mips} (3\%).

Some names point at an architecture directly such as \texttt{arm7} while others include a file ending.
Shell scripts make up 8\% (EU) vs. 6\% (US). These are platform-independent and usually perform further tasks such as downloading architecture-specific binaries.
The \texttt{mips} architecture takes a share of 9\% vs. 5\% while \texttt{x86} makes up only 5\% vs. 2\%.
This composition fits the increasing focus of malware on home gateways and IoT devices, which  could be observed over recent years.

Characterizing scanners and hosters by their origin AS shows large diversity.
Only a single AS originates about 10\% of the scans in both regions, AS~202425 (int-network).
In contrast, the hosters appear consolidated among a few ASes. 
The top four hosters make up 50\% of the events, lead by AS~16276 (ovh) with 18\% (EU) vs. 23\% (US).
AS~42864 (giganet-hu) serves 10.5\% (EU) vs. 5.4\% (US) downloads and stands out as an ISP. This ISP contributes a larger variety of names, but the most common name \texttt{le.bot.arm7} hints at a botnet. 

Two more ASes, AS~206898 (bladeservers) and AS~36352~(as-colocrossing), contribute around 10\% in both regions.
These hosters are traditional cloud providers and the overall numbers show that a small group of hosters play a large role in the distribution of (potential) malware.

\subsection{Analyzing Executables}
\label{sec:payload:downloads}

In the next step, we automatically fetch files (\ie binary files and shell scripts) to which the downloaders of the second scan phase refer.
Spoki supplies the URLs in real-time, but we rate limit the download process for repeating URLs by one download per hour.
Accessing URLs within one hour leads to success rates slightly below 50\%, presumably many scanning campaigns run outdated configurations. 
1\% of the URLs include broken hosts (\eg \texttt{YOURIPHERE}).

Over the course of 4 months in 2021 we downloaded 12,319 executable files with 250 distinct hashes (2.2\%).
While we again see a large variety of names such as {\tt arm7}, {\tt b3astmode.arm7}, or {\tt init.sh}, the most frequent name was a variant of {\tt Mozi} with more than 80\%.
Those files exhibit three different suffixes: {\tt m}, {\tt a}, and {\tt 7}.
The Mozi botnet has been around since 2019 and gained new capabilities since then~\cite{mozi-08-2021}.
Its P2P approach becomes apparent, because the IP~addresses of the hosters coincide with those of the scanners that sent the payload.

Analyzing the file types of the downloads, the majority~(70\%) are 32-bit ELF binaries (67\% MSB MIPS, 26\%~LSB~ARM, 5\% LSB MIPS, 1\% LSB Intel 80386).
In addition, we observe 14.8\% shell scripts that configure the host, download binaries of multiple architectures, and run them.
The remaining files are classified as 10.8\% ASCII text (nearly 90\% of these are shell scripts), 4\% HTML (most carry a warning for suspected phishing in the title), and two ELF 64-bit x86 binaries.

We classified the downloaded binaries using the malware database VirusTotal~\cite{virustotal}.
For each unique file, we queried VirusTotal from the first detection of the file by Spoki until its appearance in the database.
Binaries already known were considered \textit{old}, binaries later discovered by VirusTotal are \textit{new}, and binaries not appearing are considered \textit{benign}.
The classification results of our four months analysis are as follows: 18\% benign, 65\% old exploits, and 17\% new exploits.
Almost two-thirds (62.4\%) of the hashes are classified as Trojans, and 8\% are classified as downloaders.
Among them 58.8\% are labeled \textit{Mirai}.

Our data was collected from February to March and August to September 2021 with a four months pause in between.
While Mozi dominates both time frames, we can see a shift in the source countries of the Mozi hosters.
During the first period Serbia dominates with 65\% of the downloads, followed by India (21\%) and China (4.8\%).
In the second period Serbia disappeared and the origin shifted towards China, now supplying 69\%.
India dropped to 11.3\%.
This highly volatile behavior could signify a shift in infections from one region to another, but could also be the result of evolving scan strategies with new instructions for bots.

\section{Localized Two-Phase Scanning}
\label{sec:phenomenons}

We observed \one scanning of services specifically tailored to regional deployment and \two a geographic proximity between victims and origins.
In this section, we first describe our observation of locality in the telescopes, then we examine correlations between sources, targets, and the vantage point based on IXP~data.

\subsection{Scanning Locality at Telescopes}

\begin{figure} %
  \centering
  \includegraphics[width=\linewidth]{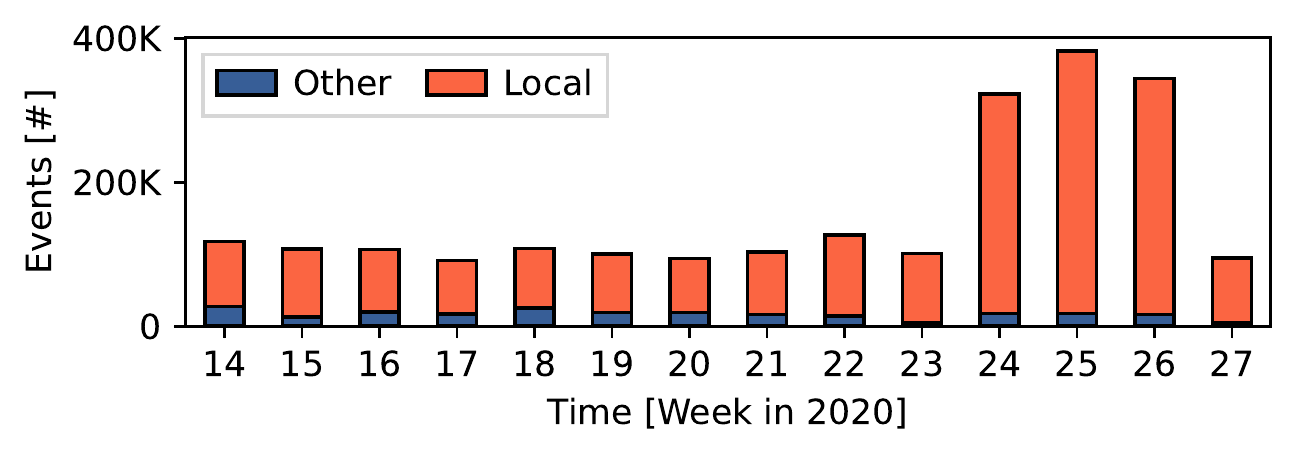}
  \caption{Share of two-phase events from the ten most contributing prefixes observed in the EU aggregated by locality. The orange bar shows the share originating from prefixes in the same /16 IPv4 address block as our /24 telescope.}
  \label{fig:phenomenons:topology}
\end{figure}

We selected the ten routable prefixes with the highest activity to look behind the scenes of scanners and estimate their scopes. These top prefixes contribute 35--40\% of all scanning events, depending on the vantage point. While examining data from the EU telescope, we found that an astonishing six out of the top ten prefixes (by count of two-phase events) are located in the same /16 block as our /24 network telescope. The shares are visualized in \autoref{fig:phenomenons:topology}.

Our observation point in the EU consists of 256 addresses in a single /24 routable prefix, which we advertise from our own autonomous system.
The covering /16 prefix splits into different networks that are distributed among other ASes.
Correspondingly, the prominent scanning activities emerge from different sub-prefixes (and providers) of this /16 prefix.
Providers are also located in different countries -- the Ukraine, Poland, and Russia.

Traffic originated from only one single address within each sub-prefix, with the exception of one case where addresses switched during our observation period with no overlap in the activity windows---a likely re-addressing of the same host.
These hosts seem not tied to specific service campaigns but scanned various service ports, \ie 8080, 80, 8082, 8081, 82, 81, 23, and 7547.

Topological locality in scanning has been previously noticed in specific contexts~\cite{cgcps-omeis-05,rb-sssim-19}, but could not be pinned down systematically. We suspect that traffic localization occurs significantly at regional level. For this reason, we widen our analysis and examine our U.S. vantage point for events originating from the related prefixes. We find no evidence of any topological correlation between scanners and this `victim' (the U.S. vantage point), which confirms that locality patterns largely depend on regions and cannot be generalized on a global scale.

\subsection{Scanning Locality at a European IXP}
\label{sec:local_scanners_ixp}

Motivated by the locality observation in our EU network telescope---\ie it receives a significant share of irregular traffic from networks of the same /16 prefix---we extend our analysis to a European IXP.
At the IXP, we can examine all (sampled) traffic flows and check whether other source prefixes emit irregular SYNs targeting the same covering /16 prefix.

We observe local, irregular SYNs for a total of 370 /16 prefixes,
which---since a /16 prefix hosts about 65k target addresses and considering a sampling rate of 1:16k---amounts to about 150 packets per host in each prefix.
These packets are largely targeted towards specific ports and match the behavior of two-phase scanners we observed.
96\% of local SYNs target telnet and router management ports (23, 7547, 8291), which notably deviates from the non-local, irregular traffic (12\% for the top three ports 80, 443, 23).
In addition, the most active sources are situated in ISP networks and in multiple cases we were able to identify consumer grade routers (MikroTik) based on Shodan information.
This indicates that infected home routers search for victims of their kind in the adjacent address space.

\begin{figure}[t]
  \centering
  \includegraphics[width=0.9\linewidth]{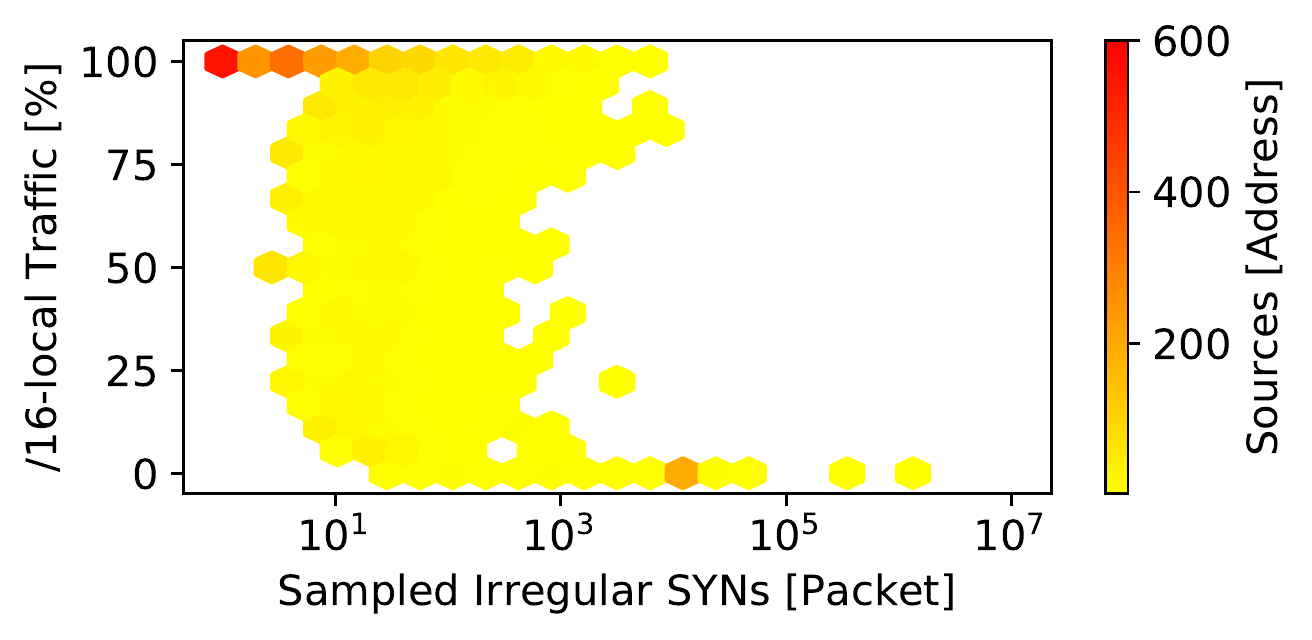}
  \caption{Share of local, irregular SYNs for scanners visible at the European IXP. The sources in the top left corner sent all of their irregular SYNs to a destination within the same /16. (Sampling rate: 1:16k)}
  \label{fig:local_scanners_ixp}
\end{figure}

For sources that sent at least one local, irregular SYN, we count the number of all irregular SYNs and calculate the share of local SYNs, see \autoref{fig:local_scanners_ixp}.
Scanning within the local /16 network is a common behavior among scanners in this region.
So much, that we find sources that only send a significant amount of local, irregular traffic (note the sampling rate of 1:16k).
Correlating the sources between the IXP and European telescope shows that 81\% of all sources for irregular, local traffic observed at the IXP have also completed a two-phase scanning event in the telescope.
This suggests that they are not limited to their /16 and occasionally stray to different prefixes, thus showing up at our vantage point.

Overall, we find highly targeted scans that operate locally and in specific prefix-ranges only, potentially aimed at certain consumer grade routers.
Repeating the analysis for the Asian ISP, however, shows no correlation of /16-local (nor /8-local) irregular SYNs.
Hence, we again conclude that the observed findings are a regional phenomenon of European networks which is analogous to the observations in our telescopes.
Our results extend previous findings~\cite{azmt-maubp-06} that botnets largely target /16 prefixes when performing localized scans.

\section{Related Work}
\label{sec:background}

\paragraph{Stateless Scanning} 
Scanrand \cite{k-paketto-02} introduced stateless scanning in 2002,
which was followed in 2004 by Unicorn~\cite{unicornscan}, and in 2013
by Masscan~\cite{masscan} and ZMap~\cite{dwh-zfiss-13}, the most
popular tool.
In this paper, we showed both that stateless scanning increased
significantly since 2014 and that many stateless scans are complemented by stateful scans (\eg Nmap) when the destination is responsive, often to explore vulnerabilities in more~detail.
Most recently, Izhikevich \etal introduced LZR~\cite{itd-liuis-21}, an advanced scanning system to identify Internet services and to reduce false positives due to middleboxes.

In contrast to prior work~\cite{gd-dcusd-20}, we cannot apply any kind of training phase to distinguish different scan implementations. %
Applying similar techniques to our data set leads to a high false
positive rate, which we can avoid due to a known feature set.
To cope with obfuscation of TCP sequence numbers, we focus on non-random header field values~\cite{gbd-ripst-16}.

\paragraph{Observing Scanners}
Bailey~\etal~\cite{bcjnw-imsdb-05} collected incoming payloads, before stateless scanning was introduced. 
Pang~\etal~\cite{pybpp-cibr-04} monitored IBR and built application-level responders.
Their results underscore the importance of analyzing unsolicited traffic to identify new breeds of malicious activities.
In contrast, Spoki does not run in kernel but in user mode, which eases maintenance.
Wustrow \etal~\cite{wkbjh-ibrr-10} excluded active measurements when looking at IBR. 

Durumeric~\etal~\cite{dbh-ivis-14} observed scanners from a darknet.
They found that $\approx$30\% of the HTTP(S) scan traffic in 2013-2014 relates to research institutions.
Our analysis of two-phase scanners clearly showed that many scans do not relate to benign activities, which increased since 2017~notably.

Richter~\etal~\cite{rb-sssim-19} used the firewall infrastructure of a global CDN to collect unsolicited traffic. Their vantage points are live CDN servers that run the firewalls, distributed across more than 1,000~ASes. 
They find that most of the observed scans do not cover the entire Internet, but remain partial, with 30\% even touching only a rather confined, local region of the address space. Unfortunately, no clear signature of IP-locality could be identified. Instead, a different target of the scanners became visible,  emphasizing vulnerable home gateways.
The measurement was purely passive and considered the vantage points as a combined set.
In our work, we showed that reacting to connection requests is crucial to reveal malicious intents and to analyze vantage points separately with respect to their topological as well as geographical distribution.
\setlength{\tabcolsep}{2pt}
\renewcommand{\arraystretch}{0.4}
\begin{table}%
  \caption{Comparison of prior work and Spoki.} %
  \label{tab:background:rw}
  \centering
  \begin{tabular}{ p{2.2cm} llll l l l l l l l l l l }
    \toprule
    & \multicolumn{7}{c}{Monitoring System}
    &
    \multicolumn{4}{c}{Key Results}
    \\
    \cmidrule(rl){2-8}
    \cmidrule(l){9-12}
    \cmidrule(){13-15}
    Measurement Year
          & \rotatebox[origin=c]{90}{Flow data}
          & \rotatebox[origin=c]{90}{Telescope}
          & \rotatebox[origin=c]{90}{Honeypot}
          & \rotatebox[origin=c]{90}{CDN}
          & \rotatebox[origin=c]{90}{LAN}
          & \rotatebox[origin=c]{90}{Active}
          & \rotatebox[origin=c]{90}{Location}
          & \rotatebox[origin=c]{90}{2-phases}
          & \rotatebox[origin=c]{90}{Ports}
          &  \rotatebox[origin=c]{90}{Topology}
          & \rotatebox[origin=c]{90}{Geography} \\
    \midrule
    
         1994-'06 
    \cite{apt-bhs-07} 
          & \xmark
          & \xmark
          & \xmark
          & \xmark
          & \cmark
            & \cmark
                & US 
                & \xmark
                            & \cmark %
                              & - 
                                & - \\
    
         '04 
    \cite{cgcps-omeis-05} 
          & \xmark
          & \cmark
          & \xmark
          & \xmark
          & \xmark
            & \xmark
                & US 
                & \xmark
                            & - %
                              & \cmark %
                                & - \\ %
    
         '04 
    \cite{pybpp-cibr-04} 
          & \xmark
          & \xmark
          & \xmark
          & \xmark
          & \cmark
            & (\cmark)
                & US 
                & \xmark
                            & \cmark
                              & \cmark  %
                                & - \\ %
    
         '06 
    \cite{azmt-maubp-06} 
          & \xmark
          & \cmark
          & \cmark
          & \xmark
          & \xmark
            & \cmark
                & US, PL%
                & \xmark
                            & - %
                              & \cmark %
                                & (\xmark) \\ %
    
         '06--'10 %
    \cite{wkbjh-ibrr-10}
          & \xmark
          & \cmark
          & \xmark
          & \xmark
          & \xmark
            & \xmark
                & US%
                & \xmark
                            & \cmark %
                              & \cmark %
                                & - \\ %
    
         '08--'15 
    \cite{bdcsk-libro-15} 
          & \cmark
          & \xmark
          & \xmark
          & \xmark
          & \xmark
            & \xmark
                & US 
                & \xmark
                            & \cmark %
                              & \cmark %
                                & - \\
    
         '11 
    \cite{dkcpp-assb-15} 
          & \cmark
          & \cmark
          & \cmark
          & \xmark
          & \xmark
            & \xmark
                & US 
                & \xmark
                            & \cmark %
                              & \xmark %
                                & (\xmark) \\ %
    
         '13-'14 
    \cite{dbh-ivis-14} 
          & \xmark
          & \xmark
          & \xmark
          & \xmark
          & \cmark
            & \xmark
                & US 
                & \xmark
                            & \cmark %
                              & - %
                                & - \\ %
    
         '15
    \cite{gbd-ripst-16}
          & \xmark
          & \cmark
          & \xmark
          & \xmark
          & \xmark
            & \xmark
                & - 
                & \xmark
                            & - %
                              & - 
                                & - \\ %
    
         '16
    \cite{wktpl-hs-18}
          & \xmark
          & \xmark
          & \xmark
          & \xmark
          & \cmark
            & \xmark
                & KR
                & \xmark
                            & \cmark %
                              & \cmark %
                                & -\\ %
    
         '18
    \cite{rb-sssim-19}
          & \xmark
          & \cmark
          & \xmark
          & \cmark
          & \xmark
            & (\cmark)
                & 156 cou.
                & \xmark
                            & - %
                              & \cmark %
                                & - \\
    
    \midrule
         '20-'21 
    Spoki 
          & \cmark
          & \cmark
          & \xmark
          & \xmark
          & \xmark
            & \cmark
                & US, EU 
                & \cmark
                            & \cmark
                              & \cmark 
                                & \cmark \\
    
    \bottomrule
  \end{tabular}
\end{table}
Blaise~\etal~\cite{bbcs-dzaup-20} found that preliminary scans target
a subset of the IPv4 space, which makes them only visible in specific
data sets.
We confirm and expand this observation in our study.

Overall, our unique measurement setup of reactive, distributed network telescopes, focusing on two-phase scanners offers new insights on scanning behavior.
We can confirm that---for some setups---scans often occur from adjacent address space.
In contrast to related work, we also found that scanners select services to target based on the location of a destination, indicating that attackers execute informed strategies.
To our surprise, origin countries of scanners are visible independently of the vantage point but the intensities of the scanning activities depend on the locations of the telescopes.
\autoref{tab:background:rw} compares most related prior work with Spoki.

\paragraph{Honeypots}
\label{sub:honeypots}
Spoki operates somewhere between pure network telescopes and honeypots~\cite{nawrocki2016survey, bringer2012survey}. 
In contrast to honeypots~\cite{p-vhf-04}, Spoki does not emulate any services or complete network stacks. It runs on a given IP prefix without knowledge of individual IP~addresses or sockets.
Spoki supports source-specific state, which we use to avoid an impact on scalability. %
Honeypots deployed in different networks observe vastly different scanning activities~\cite{whlisch2013design}, which motivated deploying Spoki in geographically and topologically distributed~networks.

Honeyd~\cite{urll-saam-17,fycts-rvhc-06} runs virtual hosts to study different servers.
In Spoki, we do not need to virtualize single hosts, which reduces overhead.
Vrable~\etal~\cite{vmcmv-sfcpv-05} present Potemkin, a virtual honeyfarm that creates VMs on demand to capture incoming traffic. Potemkin achieves scalability based on VMs that only remain active as long as needed. %
The paper presents performance and runtime insights from a 10 min. deployment of Potemkin on a /16 network. Handling the incoming traffic required 58 VMs on average with peaks into the thousands.

GQ~\cite{kwkcp-qpcmm-11} provides a platform to research malware at scale while keeping it contained. In addition to intentional infection, GQ supports a honeyfarm model where machines are exposed to the Internet. 
Flexible containment policies address the requirements for individual malware but may require manual work to fit. GQ does not focus on scalability as Potemkin~\cite{vmcmv-sfcpv-05} or Spoki do and only targets a ``moderate'' number of hosts.

Potemkin and GQ are orthogonal to Spoki. While Potemkin could
potentially fit into a similar deployment type---not as an add-on to a
telescope but as a replacement---GQ could be an additional deployment
where malware collected by Spoki could be run for further analysis.

Spoki collects packets after a TCP handshake. While packets might
include malware, Spoki does not perform any other actions aside from
identifying download instructions that use wget and curl to acquire
the linked data. In contrast, high interaction honeypots emulate full
systems and potentially allow the execution of malware.

\section{Discussion and Outlook}
\label{sec:conclusion}

In this paper, we identify and systematically explore two-phase scanning, which appears as a separate, parallel wave of malicious activities on the Internet since 2014. We developed Spoki, a highly scalable analysis tool that transforms passive telescopes into reactive measurement platforms and interacts with TCP scanners.

Two-phase scanners follow a stateless approach to initially scan for victims.
This enables them to explore the Internet rapidly and poses a particular threat to the community.
Although originally used by researchers, stateless scanning is now linked to malware.
The low technical requirements enable everyone and every device to participate in Internet-wide scans.
At such high scanning speeds attackers can identify and attack vulnerable devices leaving little time to react.
They can thus construct botnets quickly.
Our observations have practical and operational implications.

\paragraph{Security Takeaways}
Independently of the exploited protocol and vulnerability, two-phase scans act as a catalyst.
To protect against two-phase reconnaissance scans, we encourage
\emph{network operators} to deploy alert or filter rules
(\autoref{apx:ips-signatures}), while scanning for research purposes 
could be accepted.

\emph{System maintainers} should be aware that their vulnerable devices can be discovered within hours and any exploit may turn their systems into an active scanner. They should monitor for malicious activities and  shorten update cycles. 

\emph{Developers} of monitoring tools should support software that reacts to irregular TCP SYNs even though those SYNs do not comply with common TCP behavior, otherwise \emph{users} of monitoring systems, including operators and researchers, will see fewer types of attacks.
Our analysis of the transport payload underscores that lightweight transport-layer handshakes are sufficient to capture the locations of malware, which can  be fetched asynchronously (and from different locations compared to the measurement probe) for further inspection.

By correlating two \emph{reactive} network telescopes and additional vantage points in the Internet core (IXP, ISP), we unveil the differences between traditional one-phase and two-phase scanners. We find two-phase scanners to be significantly more targeted towards their victims. Repeatedly launched on home routers, two-phase scanners are also used for location-specific scans by various malware types. These locality features could be drilled down in the dimensions of physical geolocation as well as topological proximity in the IP address space, which guides \emph{network operators} where to deploy probes.
Furthermore, these results show that two-phase scanners operate more informed and explicitly target regions with vulnerable devices.
These insights help to inform \emph{policy-makers} on the need to support long-term software updates and advise for deploying protective measures.

\paragraph{Outlook}
In the future, we will use Spoki to assess the evolution of scanning through enhanced capabilities and on a larger scale. We will extend Spoki with fingerprints of multiple scan and malware families to detect and classify them automatically, thereby focusing on dedicated behavioral analysis of individual scanning campaigns.
We will also expand our telescopes to cover portions of the IPv6 address space.

\paragraph{Artifacts}
All artifacts of this paper are available at \url{https://spoki.secnow.net}.

\paragraph{Acknowledgements}
We would like to thank our shepherd Zakir Durumeric and the anonymous reviewers for their helpful feedback.
Shane Alcock is gratefully acknowledged for early discussions on this topic, as is Matthew Luckie for his work on Scamper.
We also thank Jeremy Turner for sharing his insights into malware activities.
We thank the RIPE~NCC for providing us temporary Internet ressources to conduct our measurements.
This work was partly supported by National Science Foundation grant CNS-1730661 and the German Federal Ministry of Education and Research (BMBF) within the project PRIMEnet.

\label{lastpage}

\bibliographystyle{plain}
\bibliography{cites}

\begin{thebibliography}{10}

\bibitem{azmt-maubp-06}
Moheeb Abu~Rajab, Jay Zarfoss, Fabian Monrose, and Andreas Terzis.
\newblock {A Multifaceted Approach to Understanding the Botnet Phenomenon}.
\newblock In {\em Proc. of ACM IMC}, pages 41--52, New York, NY, USA, 2006.
  ACM.

\bibitem{apt-bhs-07}
Mark Allman, Vern Paxson, and Jeff Terrell.
\newblock {A Brief History of Scanning}.
\newblock In {\em Proc. of ACM IMC}, pages 77--82, New York, NY, USA, 2007.
  ACM.

\bibitem{aabbb-umb-17}
Manos Antonakakis et~al.
\newblock {Understanding the Mirai Botnet}.
\newblock In {\em Proc. of USENIX Security Symposium}, pages 1093--1110,
  Vancouver, BC, Aug 2017. {USENIX}.

\bibitem{bcjnw-imsdb-05}
M.~Bailey, E.~Cooke, F.~Jahanian, et~al.
\newblock {The Internet Motion Sensor: A Distributed Blackhole Monitoring
  System}.
\newblock In {\em Proc. of NDSS}. Internet Society, Jan 2005.

\bibitem{bdcsk-libro-15}
K.~Benson, A.~Dainotti, K.~Claffy, A.C. Snoeren, and M.~Kallitsis.
\newblock {Leveraging Internet Background Radiation for Opportunistic Network
  Analysis}.
\newblock In {\em Proc. of ACM IMC}, pages 423--436, New York, NY, USA, 2015.
  ACM.

\bibitem{bbcs-dzaup-20}
A.~Blaise, M.~Bouet, V.~Conan, and S.~Secci.
\newblock {Detection of zero-day attacks: An unsupervised port-based approach}.
\newblock {\em Computer Networks}, 180:107391, Oct 2020.

\bibitem{bringer2012survey}
M.L. Bringer, C.A. Chelmecki, and H.~Fujinoki.
\newblock {A Survey: Recent Advances and Future Trends in Honeypot Research}.
\newblock {\em {IJCNIS}}, 4(10):63--75, Sep 2012.

\bibitem{caida-relationships}
{CAIDA UCSD}.
\newblock {AS Classification Dataset}, 2020.
\newblock \url{https://www.caida.org/data/as-classification}.

\bibitem{cgcps-omeis-05}
M.~Casado, T.~Garfinkel, W.~Cui, V.~Paxson, and S.~Savage.
\newblock {Opportunistic Measurement: Extracting Insight from Spurious
  Traffic}.
\newblock In {\em Proc. of HotNets}, New York, NY, USA, Nov 2005. ACM.

\bibitem{jspa-mdlrh-20}
Joao Ceron, Christian Scholten, Aiko Pras, and Jair Santanna.
\newblock {MikroTik Devices Landscape, Realistic Honeypots, and Automated
  Attack Classification}.
\newblock In {\em Proc. of IEEE/IFIP NOMS}, pages 1--9. IEEE, Apr 2020.

\bibitem{chs-rapc-16}
D.~Charousset, R.~Hiesgen, and T.C. Schmidt.
\newblock {Revisiting Actor Programming in C++}.
\newblock {\em Computer Languages, Systems \& Structures}, 45:105--131, Apr
  2016.

\bibitem{dbkck-eiasu-14}
A.~Dainotti, K.~Benson, A.~King, K.~Claffy, M.~Kallitsis, E.~Glatz, and
  X.~Dimitropoulos.
\newblock {Estimating Internet Address Space Usage Through Passive
  Measurements}.
\newblock {\em SIGCOMM CCR}, 44(1):42--49, Dec 2013.

\bibitem{dkcpp-assb-15}
A.~Dainotti, A.~King, K.~Claffy, F.~Papale, and A.~Pescap\`e.
\newblock {Analysis of a "/0" Stealth Scan from a Botnet}.
\newblock {\em IEEE/ACM ToN}, 23(2):341--354, Apr 2015.

\bibitem{dhjkl-dahak-07}
G.~DeCandia, D.~Hastorun, M.~Jampani, et~al.
\newblock {Dynamo: Amazon's Highly Available Key-Value Store}.
\newblock {\em SIGOPS OSR}, 41(6):205--220, Dec 2007.

\bibitem{dbh-ivis-14}
Z.~Durumeric, M.~Bailey, and J.~Halderman.
\newblock {An Internet-Wide View of Internet-Wide Scanning}.
\newblock In {\em Proc. of USENIX Security Symposium}, pages 65--78, 2014.

\bibitem{dwh-zfiss-13}
Z.~Durumeric, E.~Wustrow, and J.~Halderman.
\newblock {ZMap: Fast Internet-Wide Scanning and its Security Applications}.
\newblock In {\em Proc. of USENIX Security Symposium}, pages 605--620.
  {USENIX}, Aug 2013.

\bibitem{netacuity}
Digital Element.
\newblock {NetAcuity Edge}.
\newblock
  \url{https://www.digitalelement.com/solutions/ip-location-targeting/netacuity-edge/}.

\bibitem{fvgpn-bbtdo-21}
Álvaro Feal et~al.
\newblock {Blocklist Babel: On the Transparency and Dynamics of Open Source
  Blocklisting}.
\newblock {\em IEEE TNSM}, 18(2):1334--1349, Apr 2021.

\bibitem{suricata-website}
Open Information~Security Foundation.
\newblock {Suricata | Open Source IDS / IPS / NSM engine}.
\newblock Website, Oct 2020.
\newblock \url{https://suricata-ids.org/}.

\bibitem{fycts-rvhc-06}
Xinwen Fu et~al.
\newblock {On Recognizing Virtual Honeypots and Countermeasures}.
\newblock In {\em 2nd IEEE Intern. Symp. on Depend., Auton. and Secure Comp.},
  pages 211--218, 2006.

\bibitem{gbd-ripst-16}
V.~{Ghiette}, N.~{Blenn}, and C.~{Doerr}.
\newblock {Remote Identification of Port Scan Toolchains}.
\newblock In {\em Proc of IFIP NTMS}, pages 1--5, 2016.

\bibitem{g-meim-13}
Robert Grahahm.
\newblock {Masscan: the entire Internet in 3 minutes}.
\newblock Blog post, Errata Security, Sep 2013.
\newblock
  \url{https://blog.erratasec.com/2013/09/masscan-entire-internet-in-3-minutes.html}.

\bibitem{masscan}
Robert Graham.
\newblock {MASSCAN}, Jul 2013.
\newblock \url{https://github.com/robertdavidgraham/masscan}.

\bibitem{greynoise}
GreyNoise.
\newblock {IP Lookup API}.
\newblock \url{https://greynoise.io}.

\bibitem{gd-dcusd-20}
H.~Griffioen and C.~Doerr.
\newblock {Discovering Collaboration: Unveiling Slow, Distributed Scanners
  based on Common Header Field Patterns}.
\newblock In {\em Proc. of IEEE/IFIP NOMS}, pages 1--9, Piscataway, NJ, USA,
  Apr 2020. IEEE Press.

\bibitem{wktpl-hs-18}
Hwanjo Heo and Seungwon Shin.
\newblock {Who is Knocking on the Telnet Port: A Large-Scale Empirical Study of
  Network Scanning}.
\newblock In {\em Proc. of ASIACCS}, pages 625--636, New York, NY, USA, 2018.
  ACM.

\bibitem{itd-liuis-21}
Liz Izhikevich, Renata Teixeira, and Zakir Durumeric.
\newblock {LZR: Identifying Unexpected Internet Services}.
\newblock In {\em Proc. of USENIX Security Symposium}, pages 3111--3128.
  {USENIX} Association, Aug 2021.

\bibitem{jws-hfeda-03}
C.~Jin, H.~Wang, and K.G. Shin.
\newblock {Hop-Count Filtering: An Effective Defense against Spoofed DDoS
  Traffic}.
\newblock In {\em Proc of CCS}, pages 30--41, NY, USA, 2003. ACM.

\bibitem{jkkrs-mtuam-17}
M.~Jonker, A.~King, J.~Krupp, C.~Rossow, A.~Sperotto, and A.~Dainotti.
\newblock Millions of targets under attack: A macroscopic characterization of
  the dos ecosystem.
\newblock In {\em Proc. of ACM IMC}, pages 100--113, NY, USA, 2017.

\bibitem{k-paketto-02}
D.~Kaminsky.
\newblock {Paketto Simplified (1.0)}.
\newblock Blog post, Nov 2002.
\newblock \url{https://dankaminsky.com/2002/11/18/77/}.

\bibitem{kllpl-chrtd-97}
D.~Karger, E.~Lehman, F.~Leighton, R.~Panigrahy, M.~Levine, and D.~Lewin.
\newblock {Consistent Hashing and Random Trees: Distributed Caching Protocols
  for Relieving Hot Spots on the World Wide Web}.
\newblock In {\em Proc. of the 29th ACM STOC}, pages 654--663. ACM, May 1997.

\bibitem{kwkcp-qpcmm-11}
C.~Kreibich, N.~Weaver, C.~Kanich, W.~Cui, and V.~Paxson.
\newblock {GQ: Practical Containment for Measuring Modern Malware Systems}.
\newblock In {\em Proc. of ACM IMC}, pages 397--412, NY, USA, Nov 2011. ACM.

\bibitem{dataplane}
John Kristoff.
\newblock {DataPlane.org}, 2021.
\newblock \url{https://dataplane.org/}.

\bibitem{khrh-hhatr-14}
Marc K{\"u}hrer, Thomas Hupperich, Christian Rossow, and Thorsten Holz.
\newblock {Hell of a Handshake: Abusing {TCP} for Reflective Amplification DDoS
  Attacks}.
\newblock In {\em 8th {USENIX} {WOOT}}, San Diego, CA, Aug 2014. {USENIX}.

\bibitem{lll-iuru-05}
Robert~E. Lee, Jack~C. Louis, and Anthony de~Almeida~Lopes.
\newblock {Introducing Unicornscan. Riding the Unicorn}.
\newblock In {\em DEF CON 13},
  \url{https://www.defcon.org/html/links/dc-archives/dc-13-archive.html}, Jul
  2005.

\bibitem{lls-abiar-21}
Yassine Lemmou, Jean-Louis Lanet, and El~Mamoun Souidi.
\newblock {A behavioural in-depth analysis of ransomware infection}.
\newblock {\em IET Inf. Secur.}, 15(1):38--58, Jan 2021.

\bibitem{ldpmv-rtlca-19}
V.G. Li, M.~Dunn, P.~Pearce, D.~McCoy, G.M. Voelker, Stefan S., and Kirill L.
\newblock {Reading the Tea Leaves: A Comparative Analysis of Threat
  Intelligence}.
\newblock In {\em Proc. of USENIX Security Symposium}, pages 851--867, USA,
  2019. USENIX.

\bibitem{unicornscan}
Jack~C. Louis.
\newblock {Unicornscan}, Oct 2004.
\newblock \url{https://sourceforge.net/projects/osace/files/unicornscan/}.

\bibitem{l-ssepp-10}
Matthew Luckie.
\newblock {Scamper: A Scalable and Extensible Packet Prober for Active
  Measurement of the Internet}.
\newblock In {\em Proc. of ACM IMC}, pages 239--245, New York, NY, USA, Nov
  2010. ACM.

\bibitem{malwareworld}
MalwareWorld.
\newblock {suspiciousIPs}, 2021.
\newblock \url{https://www.malwareworld.com}.

\bibitem{njsw-fsdat-21}
Marcin Nawrocki, Mattijs Jonker, Thomas~C. Schmidt, and Matthias W{\"a}hlisch.
\newblock {The Far Side of DNS Amplification: Tracing the DDoS Attack Ecosystem
  from the Internet Core}.
\newblock In {\em Proc. of ACM Internet Measurement Conference (IMC)}, New
  York, 2021. ACM.

\bibitem{nawrocki2016survey}
Marcin Nawrocki, Matthias Wählisch, Thomas~C. Schmidt, Christian Keil, and
  Jochen Schönfelder.
\newblock {A Survey on Honeypot Software and Data Analysis}.
\newblock {\em arXiv preprint arXiv:1608.06249}, 2016.

\bibitem{mozi-08-2021}
netlab.
\newblock The mostly dead mozi and its' lingering bots, Aug 2021.
\newblock
  \url{https://blog.netlab.360.com/the-mostly-dead-mozi-and-its-lingering-bots/}.

\bibitem{pallebone}
pallebone.
\newblock {BlockIP}, 2021.
\newblock \url{https://github.com/pallebone/StrictBlockPAllebone}.

\bibitem{pybpp-cibr-04}
Ruoming Pang, Vinod Yegneswaran, Paul Barford, Vern Paxson, and Larry Peterson.
\newblock {Characteristics of Internet Background Radiation}.
\newblock In {\em Proc. of ACM IMC}, pages 27--40, New York, NY, USA, 2004.
  ACM.

\bibitem{zmap--github-scanner-go}
The~ZMap Project.
\newblock {ZGrab 2.0 Module implementation for HTTP(S).}, Jul 2020.
\newblock
  \url{https://github.com/zmap/zgrab2/blob/76d09b59c5ec1b20fcc0a172d84df99802865250/modules/http/scanner.go#L46}.

\bibitem{p-vhf-04}
Niels Provos.
\newblock {A Virtual Honeypot Framework}.
\newblock In {\em Proc. of USENIX Security Symposium}, pages 1--14, San Diego,
  CA, Aug 2004. {USENIX} Association.

\bibitem{rb-sssim-19}
Philipp Richter and Arthur Berger.
\newblock {Scanning the Scanners: Sensing the Internet from a Massively
  Distributed Network Telescope}.
\newblock In {\em Proc. of ACM IMC}, pages 144--157, New York, NY, USA, 2019.
  ACM.

\bibitem{s-dtvdo-20}
Subin Siby.
\newblock {Default TTL (Time To Live) Values of Different OS}, 2014.
\newblock \url{https://subinsb.com/default-device-ttl-values}.

\bibitem{avast-mirai-ongoing}
Avast~Software s.r.o.
\newblock {The return of the Mirai botnet}, Nov 2020.
\newblock \url{https://blog.avast.com/return-of-mirai-botnet-avast}.

\bibitem{tbakb-iiisc-20}
S.~Torabi, E.~Bou-Harb, C.~Assi, et~al.
\newblock {Inferring and Investigating IoT-Generated Scanning Campaigns
  Targeting A Large Network Telescope}.
\newblock {\em IEEE TDSC}, pages 1--17, 2020.

\bibitem{urll-saam-17}
Joni Uitto, Sampsa Rauti, Samuel Laur{\'e}n, and Ville Lepp{\"a}nen.
\newblock {A Survey on Anti-honeypot and Anti-introspection Methods}.
\newblock In {\'A}lvaro Rocha et~al., editors, {\em Recent Advances in
  Information Systems and Technologies}, pages 125--134, Cham, 2017. Springer.

\bibitem{libtrace}
New~Zealand University~of Waikato, Hamilton.
\newblock {Libtrace}, Jun 2021.
\newblock \url{https://github.com/LibtraceTeam/libtrace}.

\bibitem{urlhaus}
URLhaus.
\newblock {\url{https://urlhaus.abuse.ch/}}, 2021.

\bibitem{virustotal}
VirusTotal.
\newblock {VirusTotal}, Jan 2021.
\newblock \url{https://www.virustotal.com/}.

\bibitem{vmcmv-sfcpv-05}
M.~Vrable, J.~Ma, J.~Chen, D.~Moore, E.~Vandekieft, A.C. Snoeren, G.M. Voelker,
  and S.~Savage.
\newblock {Scalability, Fidelity, and Containment in the Potemkin Virtual
  Honeyfarm}.
\newblock In {\em Proc. of ACM SOSP}, pages 148--162, NY, USA, Oct 2005. ACM.

\bibitem{whlisch2013design}
Matthias W\"{a}hlisch, Sebastian Trapp, Christian Keil, Jochen Sch\"{o}nfelder,
  Thomas~C. Schmidt, and Jochen Schiller.
\newblock {First Insights from a Mobile Honeypot}.
\newblock {\em SIGCOMM CCR}, 42(4):305--306, Aug 2012.

\bibitem{wkbjh-ibrr-10}
E.~Wustrow, M.~Karir, M.~Bailey, F.~Jahanian, and G.~Huston.
\newblock {Internet Background Radiation Revisited}.
\newblock In {\em Proc. of ACM IMC}, pages 62--74, NY, USA, 2010. ACM.

\end{thebibliography}

\begin{appendix}
  \section{Ethical Considerations}
\label{apx:ethics-dos}

We take precautions to avoid that Spoki acts as a TCP reflector or amplifier for attacks~\cite{khrh-hhatr-14}: \one Spoki packets are small, \two~Spoki does not retransmit packets, and \three Spoki does not queue up probe requests to the same host-port combination.

\section{Two-phase Scanners Monitored at EU~Telescope}
\label{apx:scan:two-phase-sources:eu}

\begin{figure}[h!]
  \includegraphics[width=0.9\linewidth]{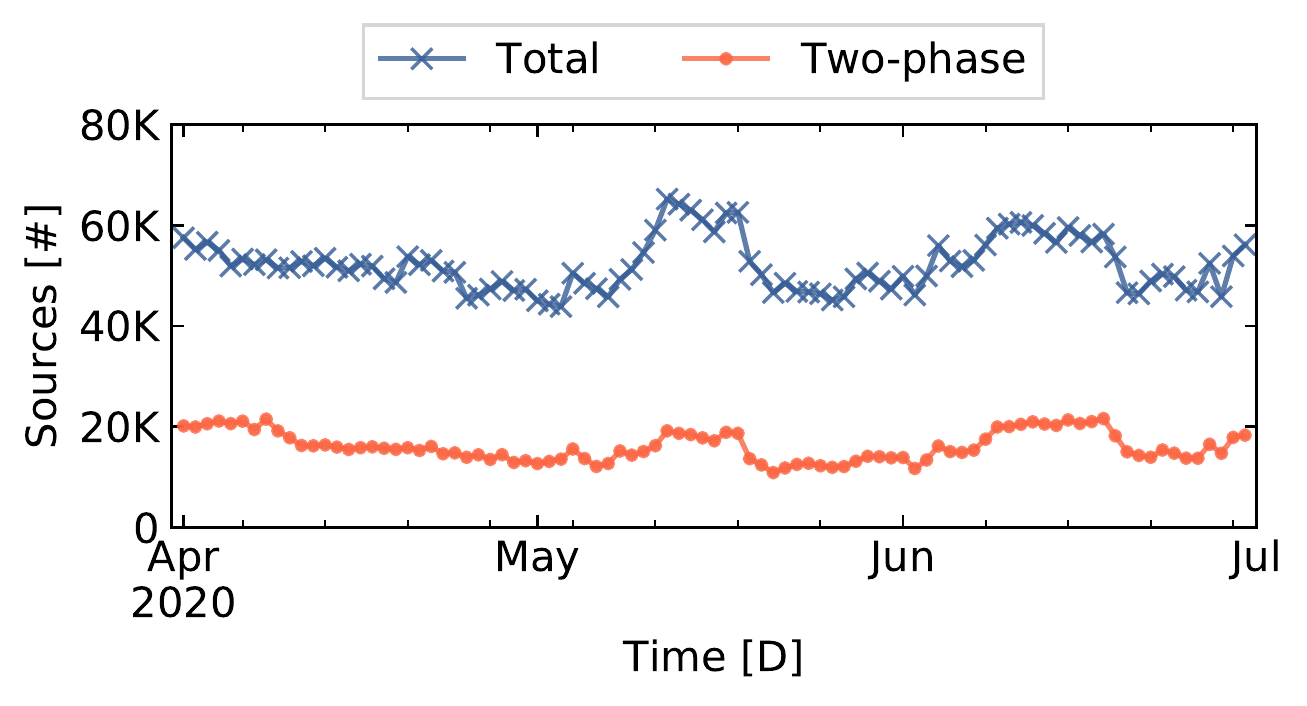}
  \caption{Share of observed two-phase sources collected in the EU.}
  \label{fig:scan:two-phase-sources:eu}
\end{figure}

We observe the same amount of sources at our Telescope in the EU compared to our telescope located in the US (compare \autoref{fig:scan:two-phase-sources:eu} with \autoref{fig:scan:two-phase-sources}).

\begin{table*}%
  \caption{Top-5 scan ports during the last 16~years, measured at our telescopes and compared to related work, which does not distinguish between two- and one-phase scanners.}
  \label{tab:eval:goals:payloads:years}
  \centering
  \begin{tabular}{ l l l l l l}
    \toprule
    \multicolumn{2}{c}{{Pre-ZMap era}}  & \multicolumn{4}{l}{{ZMap release in 2013 enabled easy two-phase scanning}} \\
    \cmidrule(lr){1-2} \cmidrule(lr){3-6}
    2004~\cite{pybpp-cibr-04} & 2010~\cite{wkbjh-ibrr-10} & 2014~\cite{dbh-ivis-14} & 2018~\cite{wktpl-hs-18} & 2020 (two-phase) & 2020 (one-phase) \\
    \midrule
    HTTP (80)                 & SMB-IP (445)              & SMB-IP (445)            & Telnet (23)             & Telnet (23)      & SMB-IP (445) \\
    NetBIOS (135)             & NetBIOS (139)             & ICMP Ping               & MS-SQL (1433)           & HTTP (81)        & Telnet (23) \\
    NetBIOS (139)             & eMule (4662)              & SSH (20)                & Netis-vuln (53413)      & MS SQL (1433)    & RDP (3389) \\
    DameWare (6129)           & HTTP (80)                 & HTTP (80)               & SSH (22)                & HTTP (7547)      & SSH (22) \\
    MyDoom (3127)             & NetBIOS (135)             & RDP (3389)              & HTTP (80)               & HTTP (80)        & Unassigned (43344) \\
    \bottomrule
  \end{tabular}
\end{table*}

\section{Evolution of Scan Ports}
\label{apx:evolution-scan-ports}
The evolution of the top-5 service scans over the last 16~years is summarized in \autoref{tab:eval:goals:payloads:years}.

\section{Integrating Spoki into a Real-Time Intrusion Prevention System}
\label{apx:ips-signatures}

In addition to our real-time measurement platform, we also contribute signatures for Suricata\cite{suricata-website}, a scalable intrusion detection system.
This integration allows to inspect the traffic of any actively used network (\ie not a telescope) for incoming two-phase scans.
Please note that these rules can be easily modified to not only \emph{alert} but also \emph{drop} such scans, which effectively protects the respective network in real-time.

\vfill\null

\begin{lstlisting}[numbers=left,frame=single,framexleftmargin=15pt,caption=Signatures for Suricata which allow for the detection of two-phase scans in actively used networks.,label=lst:apx:ids:signature]
pass tcp $HOME_NET any <- $EXTERNAL_NET any
  (msg:"IRREGULAR SYN";
  flags:S,12; ttl:>199; tcp.hdr; bsize:20;
  xbits:set,ir_syn,track ip_pair,expire 10;
  sid:1;)
  
alert tcp $HOME_NET any <- $EXTERNAL_NET any
  (msg:"TWO PHASE SCANNER";
  flags:S,12; ttl:<200; tcp.hdr; bsize:>20;
  xbits:isset,ir_syn,track ip_pair;
  sid:2;)
\end{lstlisting}

\section{Scan Origins}
\label{apx:scan-origins}
\begin{figure}[h]
  \begin{subfigure}{.49\textwidth}
    \centering
    \includegraphics[width=\linewidth]{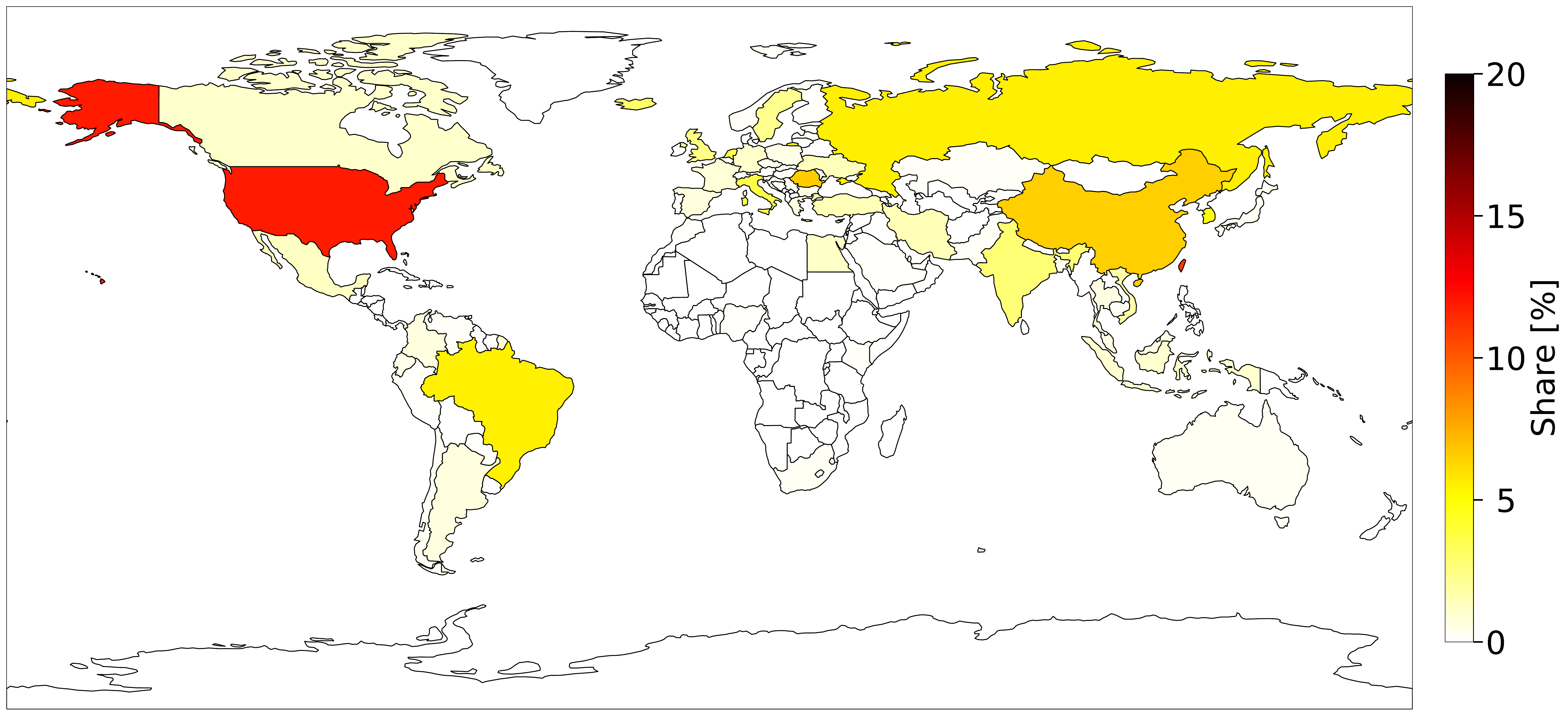}
    \caption{Collected in the US. Most of the scanners are located in the US or Taiwan, both contribute a bit more than 10\%.}
    \label{fig:eval:goals:meta:country:us}
  \end{subfigure}\hfill
  \begin{subfigure}{.49\textwidth}
    \centering
    \includegraphics[width=\linewidth]{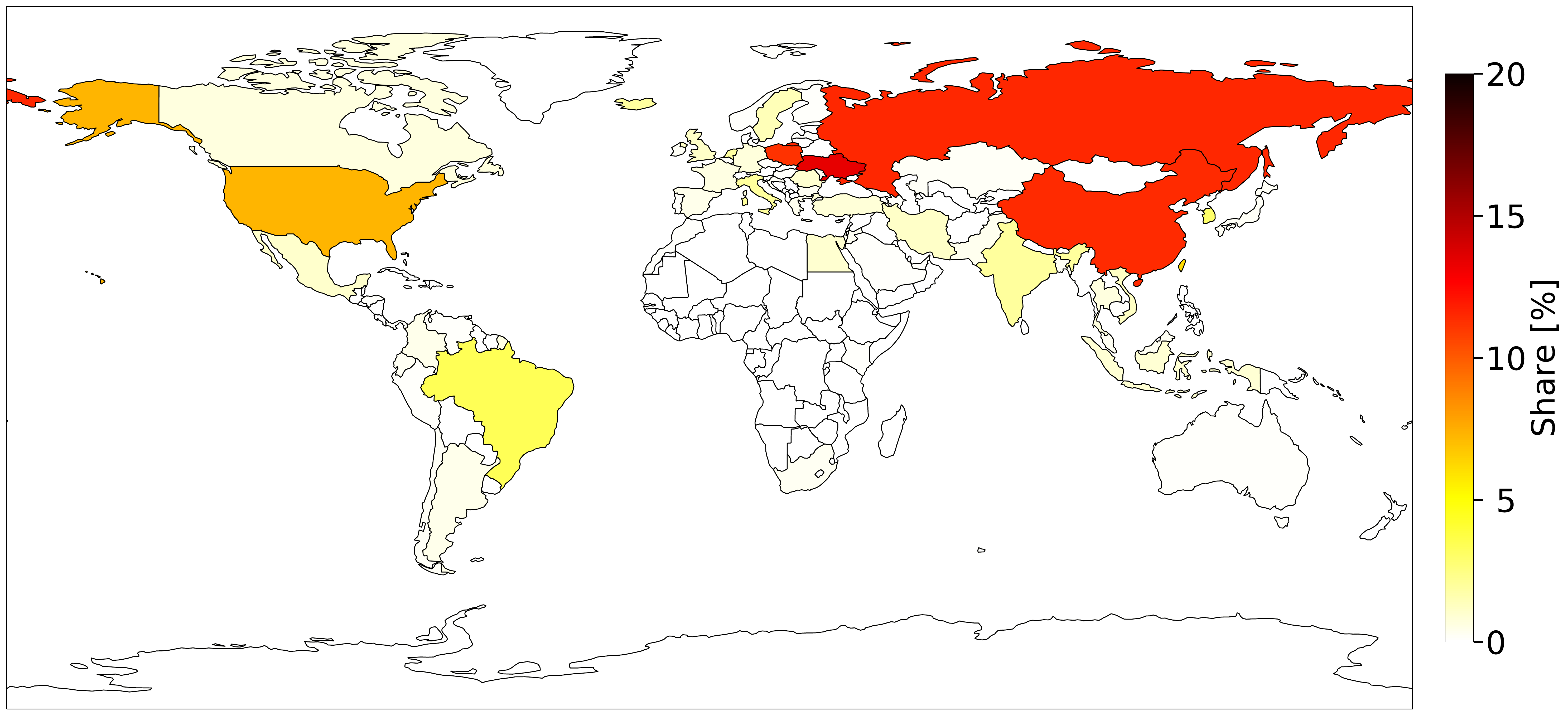}
    \caption{Collected in the EU. Most of the scanners are located in the Ukraine, Russia, China, or Poland, all contributing slightly about~10\%.}
    \label{fig:eval:goals:meta:country:eu}
  \end{subfigure}\hfill
  \caption{Share of two-phase events by country of source address. The most frequently observed origins differ by vantage point.}
  \label{fig:eval:goals:meta:country}
\end{figure}

Figure~\ref{fig:eval:goals:meta:country} visualizes the share of two-phase events by country over the period of three months, separately for each of our vantage points. Each map indicates the share of the respective events; a higher share results into a darker color. 

\end{appendix}

\end{document}